\documentclass[conference]{IEEEtran}
\IEEEoverridecommandlockouts

\usepackage{cite}
\usepackage{amsmath,amssymb,amsfonts}
\usepackage{graphicx}
\usepackage{textcomp}
\usepackage{multirow}
\usepackage{comment}
\usepackage{enumitem, array} 
\usepackage{booktabs} 
\usepackage{adjustbox}
\usepackage{caption}
\usepackage{subcaption}

\usepackage{placeins}
\usepackage{tabularx}
\usepackage{mathtools}

\usepackage[english]{babel}
\usepackage[utf8]{inputenc}
\usepackage{algorithm}
\usepackage[noend]{algpseudocode}

\newcommand{\sn}[2]{\ensuremath{{#1}\times 10^{#2}}}

\usepackage{amsthm} 
\newtheorem{theorem}{Theorem}
\newtheorem{lemma}[theorem]{Lemma}

\theoremstyle{definition}
\newtheorem{definition}{Definition}
\newtheorem{proposition}{Proposition}
\newtheorem{assumption}{Assumption}
\newtheorem{remark}{Remark}

\usepackage{lipsum}

\def\BibTeX{{\rm B\kern-.05em{\sc i\kern-.025em b}\kern-.08em
    T\kern-.1667em\lower.7ex\hbox{E}\kern-.125emX}}
\begin{document}

\title{
	Production Assessment using a Knowledge Transfer Framework and Evidence Theory }

\author{\IEEEauthorblockN{Fernando Ar\'{e}valo N.\IEEEauthorrefmark{1},
		 Christian Alison M. Piolo\IEEEauthorrefmark{4},
		 Tahasanul Ibrahim\IEEEauthorrefmark{4},
		Andreas Schwung\IEEEauthorrefmark{4}
	} 
	\IEEEauthorblockA{\IEEEauthorrefmark{1}
		Ruhr-Universit\"{a}t Bochum, Bochum 44801, Germany\\
		Email: Fernando.ArevaloNavas@ruhr-uni-bochum.de}
	\IEEEauthorblockA{\IEEEauthorrefmark{4}Department of Automation Technology\\
		South Westphalia University of Applied Sciences, Campus Soest, Germany\\
  }
  
}





\maketitle

\begin{abstract}
Operational knowledge is one of the most valuable assets in a company, as it provides a strategic advantage over competitors and ensures steady and optimal operation in machines. An (interactive) assessment system on the shop floor can optimize the process and reduce stopovers because it can provide constant valuable information regarding the machine condition to the operators. However, formalizing operational (tacit) knowledge to explicit knowledge is not an easy task. This transformation considers modeling expert knowledge, quantification of knowledge uncertainty, and validation of the acquired knowledge. This study proposes a novel approach for production assessment using a knowledge transfer framework and evidence theory to address the aforementioned challenges.  
The main contribution of this paper is a methodology for the formalization of tacit knowledge based on an extended failure mode and effect analysis for knowledge extraction, as well as the use of evidence theory for the uncertainty definition of knowledge. Moreover, this approach uses primitive recursive functions for knowledge modeling and proposes a validation strategy of the knowledge using machine data. These elements are integrated into an interactive recommendation system hosted on a backend that uses HoloLens as a visual interface. We demonstrate this approach using an industrial setup: a laboratory bulk good system. The results yield interesting insights, including the knowledge validation, uncertainty behavior of knowledge, and interactive troubleshooting for the machine operator.
\end{abstract}



\section{Introduction}\label{section__introduction}
Sustaining operational know-how guarantees companies an advantage over competitors. This can be achieved by establishing best practices that ensure the optimal operation of machines and recording troubleshooting approaches that reduce downtime  \cite{Omotayo2015}. 
This knowledge has been accumulated in company logs over the years. In the best of cases, it is recorded in the form of best-practice manuals, maintenance documents, and troubleshooting guides, so that future machine operators can access it. This type of knowledge is referred to as explicit knowledge. In comparison, tacit or implicit knowledge refers to the empirical expertise gained by operators on the shop floor. The transformation of implicit knowledge into explicit knowledge is not an easy task. The reasons for this involve lack of adequate knowledge transfer strategies, procedures and tools for institutional knowledge internalization \cite{Newell2005}. 
Knowledge transfer has been addressed through different strategies, such as peer-to-peer communication, producing written sources (e.g., books and user guides), audiovisual guides, or immersive augmented reality (AR) and virtual reality (VR) applications.
However, some of these strategies might introduce bias in the acquired knowledge \cite{Stammers2018} (e.g., in peer-to-peer communication where the sender chooses the information shared in terms of perceived relevance). Additional challenges include the quantification of knowledge uncertainty, and effective strategies to validate the extracted knowledge.

\begin{table}[!htbp]
	\centering
	\caption{List of Symbols and Abbreviations}
	\begin{tabular}{l l}
		\toprule
		\textbf{Symbol} & \textbf{Description} \\ 		
		\midrule
		$R$             & rule\\ 
		$P$             & operand  \\ 
		$O$             & operator  \\
		$C$             & condition  \\
		$m$             & mass function  \\
		$U$             & uncertainty  \\
		$w_{R}$         & rule weight  \\
		$w_{R_{a}}$     & accumulated rule weight\\
		$TU$            & knowledge tuple\\
		$SP$            & sub-process\\
		$FM$            & failure mode\\
		$C$             & causes\\
		$E$             & effects\\
		$RE$            & recommendations\\
		$k$             & sensitivity to zero factor\\
		$F$             & approximation factor\\
        $V$             & variables\\
	    $T$             & thresholds\\
		$K_{V}$         & knowledge validation\\
		$N_{V}$         & number of validation samples\\
		FMEA            & Failure Mode and Effects Analysis\\
		DSET            & Dempster Shafer Evidence Theory\\
		KPI             & Key Performance Indicator\\
		KLAFATE         & KnowLedge trAnsfer FrAmework using evidence ThEory\\
		BGS             & Bulk Good System\\
		OPC-UA          & OPC Unified Architecture\\
		MQTT            & Message Queuing Telemetry Transport\\
		AR              & Augmented Reality\\
		\bottomrule
	\end{tabular}
	\label{table__list_symbols} 
\end{table}

Knowledge transfer involves different stages such as extraction, modeling, uncertainty definition, validation, and visualization \cite{DeWit-deVries2019} \cite{Korteling2021} \cite{Anand2020} \cite{Hoffmann2019}. Thus, defining a knowledge transfer framework would provide a substantial step towards interactive knowledge transfer as it would clearly identify each stage and the relevant challenges to be addressed.
A knowledge transfer framework would allow the deployment of an interactive assessment system on the shop floor, which would provide a constant flow of valuable information concerning machine conditions to the operators, as well as a set of recommendations geared at solving process issues. 

This article proposes a novel methodology for production assessment based on an interactive knowledge transfer framework in which evidence theory is an intrinsic part. It provides a knowledge transfer framework that considers all knowledge stages, as it allows the extraction and transfer of knowledge, and is facilitated through a user interface (UI). This research identifies the existing challenges for each step in the knowledge chain and methodologically addresses them.

The contributions of this paper are summarized as follows:
\begin{itemize}

    \item Defining a methodology for the formalization of tacit knowledge based on an extended failure mode and effects analysis (FMEA) to extract knowledge from an expert panel objectively and systematically. In addition, the Dempster-Shafer evidence theory was used to evaluate the existing uncertainty factors in the extracted knowledge.
    
    \item Presenting the use of primitive recursive functions to create a knowledge-based model. This model integrates the knowledge extracted from an extended FMEA and the uncertainty defined through evidence theory.
    
    \item Presenting a strategy for knowledge validation based on key performance indicator (KPI) analysis. The KPIs are calculated using machine data in short- and long-term periods to consider the effects of knowledge across time.
	
	\item Finally, defining a strategy to embed a knowledge transfer framework into an interactive assessment system hosted in a backend. The assessment system uses an augmented reality device as the UI to enhance user experience. Moreover, we demonstrate this approach using a small-scale industrial setup.  

\end{itemize}

This article is organized as follows. In Section \ref{section__related}, we present state-of-the-art knowledge transfer stages, quantification of knowledge uncertainty, and production assessment applications. 
Section \ref{section__proposed} introduces the proposed model and methodology to address the knowledge chain. Section \ref{section_usecase} describes a practical implementation of this approach in an industrial setup. Finally, Section \ref{section_conclusions} presents the findings and remarks of the authors and provides an outlook for further research.

\section{Related Work}\label{section__related}
\begin{figure}[!htbp]
	\centering
	\includegraphics[width=0.4\textwidth]{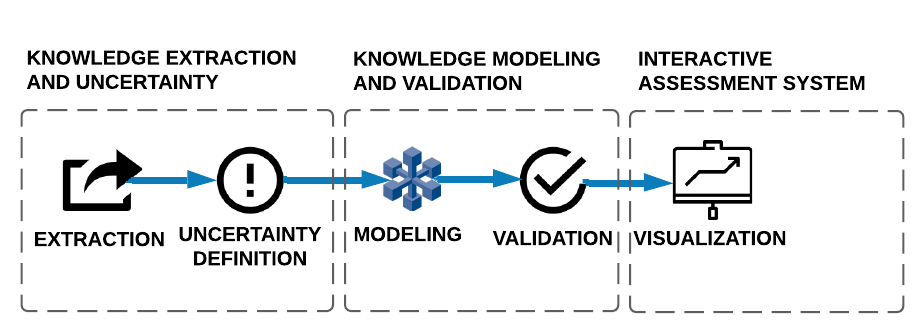}
	\caption{Knowledge Stages}\label{figure__knowledge_stages}
\end{figure}

This article presents an approach to production assessment using an interactive knowledge transfer framework. The knowledge transfer framework addresses three main points: the formalization of tacit knowledge into explicit knowledge, quantification of knowledge uncertainty, and definition of a strategy to embed the knowledge framework into an interactive assessment system. Fig. \ref{figure__knowledge_stages} shows the stages of the knowledge transfer framework.

\subsection{Extraction of Tacit Knowledge and Uncertainty Quantification}
The extraction of tacit knowledge involves a \textit{knowledge extraction} method. For this purpose, different methodologies have been proposed, which include, among others, the use of question-answering systems \cite{Agarwal2019} \cite{Su2019} \cite{Qiu2020}, the use of process mapping \cite{White2016}, ontologies \cite{Rodriguez2018} \cite{Petnga2016}, and lean manufacturing tools, such as failure mode and effects analysis (FMEA) \cite{Westermeier2013}\cite{Pytlik2016}.
However, knowledge extraction can be challenging, because the information should be representative and free of bias. Moreover, the knowledge extraction process can be time-consuming (e.g., especially for question-answering systems and interviews), or in the case of ontologies that require a high level of detail and implementation. 
Therefore, we require a methodology that allows us to store information systematically, frequently, and without bias \cite{Ji2010} to provide the machine operator with an objective, representative, and relevant assessment. Taking a systematic approach would allow overcoming information bias, as it would avoid subjective preferences in one expert’s opinion. 

When establishing a body of explicit knowledge, one crucial feature is the \textit{uncertainty quantification of the knowledge} or confidence level concerning expert statements. Knowledge inaccuracies can be attributed to slightly detailed information or the existence of new unknown events. Knowledge uncertainty has been represented using fuzzy systems \cite{Koyuncu2005}, evidence theory \cite{Kukulies2018} \cite{Wang2019a} \cite{Zhang2019}, or hybrid systems \cite{Zhang2019a}\cite{Jiang2019}.
The next step in consolidating acquired knowledge is to define a strategy for knowledge uncertainty quantification.  

\subsection{Knowledge Modeling and Validation}
\textit{Knowledge modeling} has been addressed using (rule-based) expert systems\cite{Ali2018}\cite{GomezReynoso2013}, ontologies \cite{Rodriguez2018}\cite{Petnga2016}, fuzzy systems \cite{Koyuncu2005}, and knowledge graphs \cite{Chen2020}, among others. However, modeling the extracted knowledge can be an exhausting task because of the detailed level of information to be provided, especially for expert systems and ontologies. The upside of such models is the inclusion of the expert domain in modeling, whereas the downsides are the risk of biased information, increased time consumption owing to the number of rules to be defined and the subjectivity. The maintenance of the knowledge-based model needs to be considered as the knowledge would be modified frequently.

To this end, a knowledge extraction methodology using FMEA was presented in \cite{Arevalo2019}, where the authors proposed a strategy to formalize tacit knowledge, specifically machine faults, into a knowledge-based model in a systematic manner. This knowledge-based model contains rules that can be triggered using machine data. However, although the authors proposed a strategy to digitize the knowledge quickly, and although the recommendation system detected the faults in the process, no uncertainty quantification of the knowledge rules that could assess the certainty of the triggered fault to the machine operator was provided.

\textit{Knowledge validation} is necessary to evaluate the effectiveness of the knowledge models. An evaluation performed using machine data can provide objective insights based on machine performance. Key performance indicators (KPI) have been used to assess the performance of machines and processes, such as acceptance rate, mean downtime, and operating time \cite{Meier2013}. Lindenberg et al. \cite{Lindberg2015} stressed the importance of KPIs in performance monitoring in the industry because they can help identify poor performance and, thus, create improvement potential. Meier et al. \cite{Meier2013} explained the role of KPI within the assessment of the delivery of industrial service.

\subsection{Interactive Assessment Systems}
Interactive assessment systems have industrial applications using AR for remote maintenance \cite{Mourtzis2017}, production and quality monitoring on the shop floor \cite{Segovia2015}, and cross-platform dashboards for assembly operations \cite{Cheng2013}. Written documentation is popular on the shop floor while performing troubleshooting; however, the operator must find the proper terms that identify the problem and then search for a suitable solution \cite{Mohr2015}. Online documentation eases this problem when available. However, the assessment provided to the operator is a set of statements rather than a list of recommendations with an associated confidence level. An interactive recommendation system would support the operator in the decision-making process on the shop floor. Segovia et al. \cite{Segovia2015} investigated the effectiveness of an AR-based interactive system to decrease defects on the shop floor, where the AR implementation assisted in the improvement of quality reporting and decision making while displaying necessary information to the user. Additionally, Hoffmann et al. \cite{Hoffmann2019} demonstrated the effectiveness of using the AR device HoloLens, as a tool in a cyber-physical system (CPS) for knowledge and expertise sharing in manufacturing. This study discussed the importance of visualization and interaction of knowledge transfer between knowledgeable persons and knowledge seekers through CPS. Thus, gamification positively contributes to the knowledge transfer process. 
Mourtzis et al. \cite{Mourtzis2017} presented another method for interactive assessment using AR remote assistance. The user would be able to contact remote experts for recommendations that were presented through AR scenes. This implementation managed to reduce travel costs and downtime. Kokkas et al.\cite{Kokkas2019} used holograms in an AR application to test new plant layouts as a method for interactive assessments. This paper stresses the importance of having a real environment, stating that it allows a realistic assessment of solutions based on a quantitative and qualitative approach.

Our approach is distinct from the known state-of-the-art approaches in three ways. First, we propose a holistic approach to managing the overall knowledge chain. Specifically, we concentrate on the quantification of knowledge uncertainty, as well as its inclusion in a knowledge-based model based on primitive recursive functions. Second, we propose a strategy for embedding a knowledge transfer framework into an interactive assessment system hosted in the backend. Third, we demonstrated the industrial plausibility of this approach using an industrial laboratory testbed that is comparable to industrial setups.

\section{KLAFATE: Knowledge Transfer Framework and Evidence Theory}\label{section__proposed}

\begin{figure*}[!htpb]
	\centering
	\includegraphics[width=0.8\textwidth]{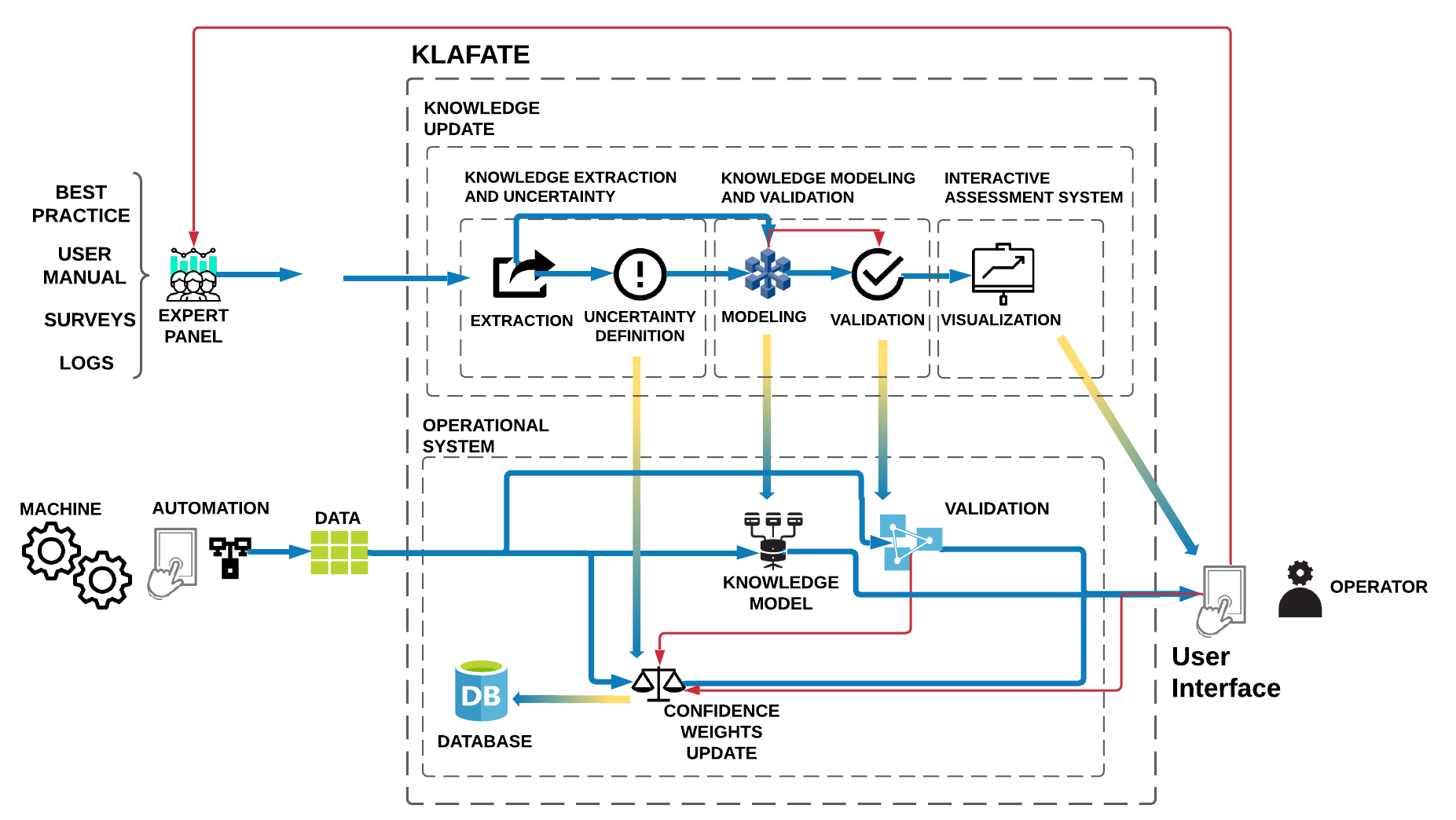}
	\caption{Knowledge Framework Overview}\label{figure__proposed_knowledge_chain}
\end{figure*}

This research proposes a user-centered approach to gather process expertise from the shop floor using a KnowLedge trAnsfer FrAmework using evidence ThEory (KLAFATE). 
The system architecture is shown in Figure \ref{figure__proposed_knowledge_chain}, which portrays the knowledge flow from its tacit form to an explicit (digitized) version. 
KLAFATE comprises two major sections: knowledge update and the operational system. 

The knowledge update section considers all necessary steps to acquire knowledge from the first time, as well as every time the knowledge needs to be updated. 
The stages of knowledge update are summarized as follows:

\begin{itemize}
    \item \textbf{Extraction of tacit knowledge and uncertainty quantification}: The \textit{tacit knowledge} from process assets is transformed into an explicit form using an extended version of the causal method \textit{failure mode and effect analysis (FMEA)}. The knowledge was \textit{extracted} from an expert panel and written in the templates of the extended FMEA. The expert panel quantified knowledge uncertainty by defining weights for each rule. Thus, each rule weight is a function of predefined criteria:
    $w_{R} = f(c_{1}, c_{2}, c_{3},...,c_{n})$, where $c_{1}$, $c_{2}$... symbolize the criteria.
    
    \item \textbf{Knowledge modeling and validation}: Knowledge rules are transformed into a \textit{knowledge model} using primitive recursive functions. Thus, the system can be represented as a switch case where each case is a knowledge rule. The knowledge rules need to be \textit{validated} regularly according to the criteria. The criteria consider a data-based method that uses the KPIs of the system to validate the rule. 
    
    \item \textbf{Interactive assessment systems}: The system interacts with the user through a \textit{visual interface}. This interface displays the assessment while experimenting a fault and allows the operator to input feedback to the system in terms of system usability. 
    \end{itemize}

The operational system receives artifacts from the knowledge update section, namely, the knowledge model, rule weights, knowledge validation strategy, and AR application. 

\subsection{Theoretical Background}

\subsubsection{Boolean Logic Rules}
A rule can be written as
\begin{equation} \label{equation__model_1}
R_{i} =  (\ P_{j}\ O_{j}\ P_{j+1}\ O_{j+1}\ ...\ P_{M-1}\ O_{M-1}\ P_{M}\ ) 
\end{equation}
where $R_{i}$ is the $i_{th}$ knowledge rule, $P_{j}$ is the $j_{th}$ operand, $O_{j}$ is the $j_{th}$ operator, and $i,j,M \in \mathbb{N}$. The operator $O_{j}$ is a logic operator (e.g., $\leq, \geq, \neq, =, \lor, \land, \neg, etc. $).

The $j_{th}$ operand $P_{j}$ can be represented as a function of process variables $V$ and process thresholds $T$ as follows:
\begin{equation} \label{equation__model_2}
P_{j}=f(V,T)
\end{equation}

Successively, the operand $ P_{j}$ can be represented using:  
\begin{equation} \label{equation__model_3}
 P_{j} = (\ P_{k}\ O_{k}\ P_{k+1}\ O_{k+1}\ ...\ P_{N-1}\ O_{N-1}\ P_{N}\ ) 
\end{equation}
where $P_{j}$ is the $j_{th}$ operand, $P_{k}$ the $k_{th}$ operand, and $O_{k}$ the $k_{th}$ operator, and $j,k,N \in \mathbb{N}$.

Thus, the operator $P_{j}$ could take one of the following forms:
\begin{equation} \label{equation__model_5}
     P_{j}=\begin{cases}
         C_{k} \\
         \neg \quad C_{k} \\
         P_{k} \quad O_{k} \quad P_{k+1}\\
         \end{cases}
\end{equation}
where $C_{k}$ is a condition that is a function of the process variables $V$ and process thresholds $T$:

\begin{equation} \label{equation__model_5a}
C_{k}=f(V,T)
\end{equation}

Thus, the condition $C_{k}$ could take one of the following forms:
\begin{equation} \label{equation__model_6}
     C_{k}=\begin{cases}
          V_{i}\ >\ T_{i}\\
          V_{i}\ ==\ T_{i} \\
          T_{i}\quad \text{if}\quad C_{i}\quad \text{else}\quad T_{i+1}
         \end{cases}
\end{equation}

The knowledge rules return a Boolean output, which signalize that a knowledge rule is active (e.g., the first two examples in Equation (\ref{equation__model_6}) return a Boolean output). In the case of the rule weights, the output is a real number in the range [0,1] (e.g., in the third example in Equation (\ref{equation__model_6}), an \textit{if-else} statement returns a real number).  

\subsubsection{Dempster Shafer Evidence Theory}

\begin{definition}[Dempster-Shafer \cite{Shafer1976}]\label{definition__dempster__1}
Let $\Theta$ be a frame of discernment, in which each focal element represents a condition. A basic probability assignment (BPA) can be defined using a function m: $2^{\theta} \rightarrow [0,1]$, whenever:
\begin{equation} \label{equation__DSET_1}
\begin{split}
m(\phi) & = 0
\end{split}
\end{equation}
\begin{equation} \label{equation__DSET_2}
\begin{split}
\sum_{A \subseteq \Theta} m(A) & = 1
\end{split}
\end{equation}
Thus, considering a frame of discernment $\Theta = \{A,B\}$, the power set $2^{\theta}$ is represented by:

\begin{equation} \label{equation__DSET_2a}
\begin{split}
2^{\theta} = \{ \phi, \{A\},\{B\}, \Theta \} \}
\end{split}
\end{equation}

The sum of the BPAs from (\ref{equation__DSET_2}) can be transformed into:
\begin{equation} \label{equation__DSET_3}
\begin{split}
S_{bpa} = \sum_{A \subseteq \Theta} m(A) & = \sum_{j=1}^{n} m_{j} = 1
\end{split}
\end{equation}

where $m_{j}$ is the $j_{th}$ focal element of $\Theta$, and j, n $\in \mathbb{N}$.
\end{definition}

The elements of $\Theta$ are considered mutually exclusive. For example, given a $\Theta = \{A,B\}$, a combination of focal elements is not possible:
\begin{equation} \label{equation__DSET_6}
\begin{split}
A \cap B = \phi
\end{split}
\end{equation}

A BPA describes the certainty of each focal element (e.g., a condition, a fault). 
Considering the weights of each focal element can help while quantifying the overall uncertainty. 
To this end, this paper presents a new weighted $S_{bpa}$, denoted as $S_{wbpa}$, that describes the overall uncertainty of a BPA by using the weights of each BPA.

\begin{proposition}\label{definition__dempster__2}
The sum of BPAs from Equation (\ref{equation__DSET_3}) can be transformed into:
\begin{equation} \label{equation__DSET_4}
\begin{split}
S_{wbpa} & = \sum_{j=1}^{n} m_{j}*w_{m_{j}} + U = 1\\
\end{split}
\end{equation}
where $w_{m_{j}}$ is the $j_{th}$ confidence weight of the BPA $m_{i}$, and $U$ is the overall uncertainty. The confidence weight $w_{m_{j}}$ represents the confidence level of the evidence $m_{i}$, which can be quantified using a predefined criteria.
\end{proposition}

The overall uncertainty of the body of knowledge can be represented as:
\begin{equation} \label{equation__DSET_5}
\begin{split}
U & = 1- \sum_{j=1}^{n} m_{j}*w_{m_{j}}\\
\end{split}
\end{equation}
where the value of $U$ will increase as the confidence weights of the focal elements of $\Theta$ decrease. Thus, a large value of $U$ corresponds to a high uncertainty in the body of knowledge. In this sense, the overall uncertainty $U$ represents the amount of unknown information or the lack of evidence.

At least one of the focal elements of $\Theta$ is different from zero: 
\begin{equation} \label{equation__DSET_8}
\begin{split}
 \forall m_{j}. \quad  m_{j}> 0
\end{split}
\end{equation}

\begin{definition}\label{definition__dempster__5}
Each confidence weight $w_{m_{j}}$ is bounded:
\begin{equation} \label{equation__DSET_7}
\begin{split}
w_{m_{j}} \rightarrow [0,1]
\end{split}
\end{equation}
\end{definition}

Knowing the value of the overall uncertainty, we could assess the confidence in the available evidence. Therefore, Proposition \ref{definition__dempster__2} paves the way to obtain an overall uncertainty measurement considering the confidence weight of each piece of evidence. However, the integrity of Equation (\ref{equation__DSET_4}) (sum of BPAs) should be preserved. For this reason, Proposition \ref{definition__dempster__2} must be consistent with Equation (\ref{equation__DSET_4}) by mathematical proof.   


\begin{lemma}\label{lemma__dempster__1}
Denote the $S_{bpa}$ and $S_{wbpa}$ as the BPA sum and BPA weighted sum with an explicit overall uncertainty definition, respectively. Then, it holds:
\begin{equation} \label{equation__DSET_5a}
\begin{split}
S_{bpa} = S_{wbpa}\\
\end{split}
\end{equation}
\end{lemma}

\renewcommand\qedsymbol{$\blacksquare$}

\begin{proof}
Considering each weight $w_{m_{j}} \rightarrow 1$, then $U= 1- \sum_{j=1}^{n} m_{j}$, and if Equation (\ref{equation__DSET_3}) holds, $U=0$. Hence, (\ref{equation__DSET_3}) is fulfilled as both sides equals to one. Likewise, considering each weight $w_{m_{j}} \rightarrow 0$, then the term $\sum_{j=1}^{n} m_{j}*w_{m_{j}}$ tends to zero, and $U=1$, thus, both sides equal to one. The first scenario represents a total certainty on the provided evidence, which in turn result in $U=0$. Whereas the second scenario represents a total uncertainty on the provided evidence, which results in $U=1$. Any other case in which $w_{m_{j}} \rightarrow [0,1]$ will result on the condition equal to one, due to the mutual cancellation of $\sum_{j=1}^{n} m_{j}*w_{m_{j}}$.  
\end{proof}

\subsection{Extraction of Tacit Knowledge and Uncertainty Quantification}

\subsubsection{Knowledge Extraction}
The lean manufacturing tool \textit{failure mode and effect analysis (FMEA)} is extended for use as a causal method to transfer tacit knowledge from the shop floor into an explicit form, which can be easily modeled as knowledge rules. The FMEA is built by an expert panel from the process, and identifies the failure modes, possible causes, and recommendations from a determined system. This research uses an extended FMEA to extract the knowledge into a digital format.

The knowledge tuple $TU$ has the form:
 \begin{equation} \label{equation__proposal__1}
\begin{split}
TU_{i} = (P, SP, FM, C, E, RE, R, w_{R})
\end{split}
\end{equation}
where, $i \in \mathbb{N}$. Each knowledge tuple has only one associated rule $R$ and only one rule weight $w_{R}$. Each failure mode $FM$ is associated with one process $P$ and one sub-process $SP$. Each $FM$ can have several causes $C$, effects $E$, and recommendations $RE$. 
Knowledge rule $R$ can be used for process optimization or troubleshooting purposes. 
This article proposes an improved version of the extended FMEA from \cite{Arevalo2019}, which consists of a spreadsheet with four templates: settings, weight update, system and component. Moreover, this study formalizes the previous approach mathematically, allowing further improvements and modifications.  

\begin{table*}[!htpb]
	\centering
	\caption{FMEA Template}\label{table__traditional_fmea}
	\includegraphics[width=0.97\textwidth,keepaspectratio]{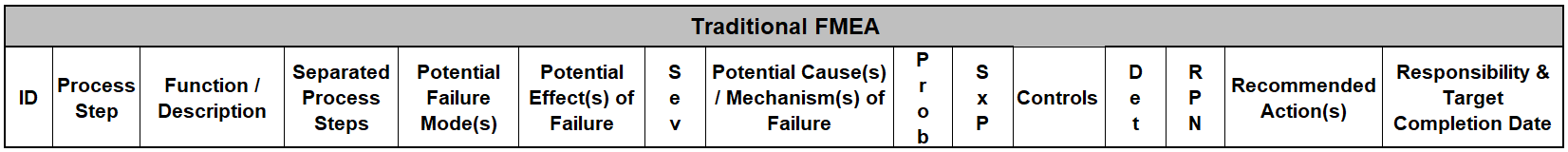}
\end{table*}

\begin{table*}[!htpb]
	\centering
	\caption{Extended FMEA Template}\label{table__extended_fmea}
	\includegraphics[width=0.87\textwidth,keepaspectratio]{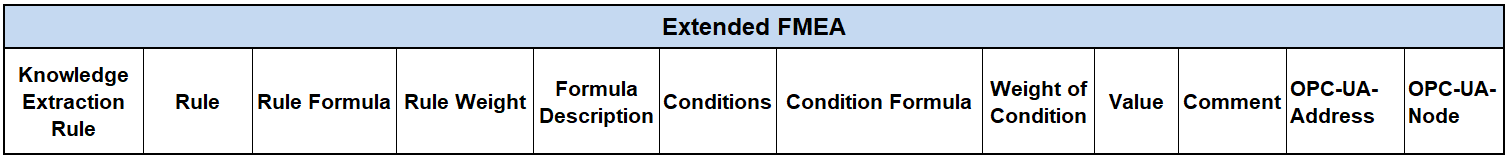}
\end{table*}

\begin{table*}[!htpb]
	\centering
	\caption{Settings Template}\label{table__fmea_settings}
	\includegraphics[width=0.57\textwidth,keepaspectratio]{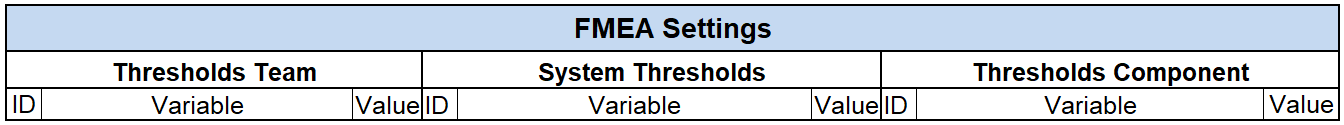}
\end{table*}


The extended FMEA provides a framework for establishing knowledge rules using templates. These rules are used to detect faults in the system, and they follow the criteria defined in the \textit{Controls - Diagnosis} of the traditional FMEA. The rules were written in a programming-friendly manner to make information parsing manageable. Each rule contains a formula for detecting a failure mode, which is a function of process variables $V$ and process thresholds $T$. This formula can be more detailed using sub-rules. 

\begin{figure*}[!htpb]
	\centering
	\includegraphics[width=0.97\textwidth,keepaspectratio]{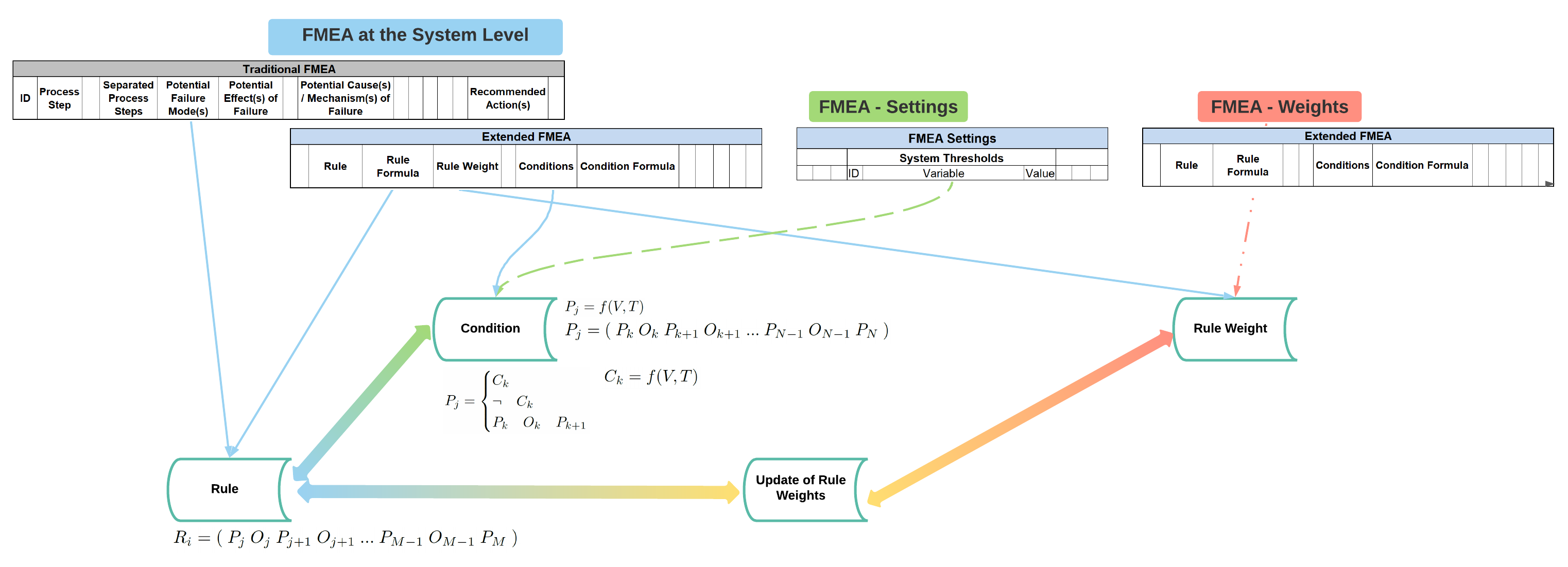}
	\caption{Relationship between templates}\label{figure__templates_relation}
\end{figure*}

The previous procedure is illustrated graphically in Figure \ref{figure__templates_relation}, which shows the relationships between the templates.


The template \textit{settings} included the thresholds and system set points in one place (see Table \ref{table__fmea_settings}). This is implemented such that the variables can be changed easily without the hassle of changing individual variables in different templates. This template included a section for the team, system, and component. Team thresholds change variables of the template \textit{weight update}, whereas the system thresholds and component thresholds affect the templates \textit{system} and \textit{component}, respectively.   
The template \textit{weight update} contains information to quantify the uncertainty of the knowledge rules as confidence weights. The confidence weights used the criteria specified by the expert panel.

\subsubsection{Uncertainty Quantification of the knowledge rules}
The uncertainty of the knowledge rules can be quantified using confidence weights for each rule. This implies finding criterion that can represent the uncertainty of the rule so that the weight for the rule can be defined. The weight of the knowledge rule will have a value in the range $[0,1]$, and it was used to assess the certainty of the rules.
The weight of each rule $w_{R}$ is built using: 
\begin{equation} \label{equation__weight__1}
\begin{split}
w_{R_{j}} = \frac{1}{N_{R}}\sum_{i=1}^{N_{R}} w_{R_{C_{i}}}(V,T)
\end{split}
\end{equation}
where $V$ is the process variable, $T$ is the process threshold, $w_{R}$ is the rule weight, $w_{R_{C_{i}}}$ is the $i_{th}$ criterion for the rule weight, and $N_{R} \in \mathbb{N}$. 
The criteria for the confidence weights of the knowledge rules are found in the template \textit{weight update}. It is important to note that the expert panel can define the extent of the rule criteria, so these criteria could be composed of one or several criteria.
Each weight $w_{R}$ can contain a $N_{R}$ number of sub-weights $w_{R_{C_{i}}}$. The expert panel defines the criteria to conform to each of these $w_{R_{C_{i}}}$, which is a function of the variables $V$ and thresholds $T$ in the template.  
However, this research uses three main criteria to conform to the weight of a rule: the weight of the expert panel $w_{R_{C_{1}}} = w_{P}$, weight of the KPI compliance $w_{R_{C_{2}}} = w_{K}$, and weight of the User Rating $w_{R_{C_{3}}} = w_{U}$.
Once the system is in operation, the confidence weights can be updated dynamically using the historical values. The accumulated value of the weight $w_{R_{a}}$ can be calculated as:
\begin{equation} \label{equation__weight__2}
\begin{split}
w_{R_{a}} = \frac{1}{N_{R_{a}}}\sum_{j=1}^{N_{R_{a}}} w_{R_{j}}
\end{split}
\end{equation}
The \textit{weight of the expert panel} $w_{P}$ is defined using:
\begin{equation} \label{equation__weight__3}
\begin{split}
w_{R_{C_{1}}} & = w_{P}(V_{t},T_{t}) \\ 
& = \frac{1}{N_{P}}\sum_{i=1}^{N_{P}} w_{M_{i}}
\end{split}
\end{equation}
, where $N_{R}, N_{R_{A}} \in \mathbb{N}$, and $V_{t}$ represents the team variables, $T_{t}$ the team thresholds from the template \textit{settings}, and $w_{M_{i}}$ represents the weight of the $i_{th}$ member of the expert panel.  

The weight of the expert panel $w_{M_{i}}$ is defined using:
\begin{equation} \label{equation__weight__4}
\begin{split}
w_{M_{i}} = \frac{1}{N_{M}}\sum_{j=1}^{N_{M}} w_{M_{C_{j}}}
\end{split}
\end{equation}
where $N_{M}, j, i \in \mathbb{N}$, and $w_{M_{C_{j}}}$ is the weight of the $j_{th}$ criteria $C_{j}$ to evaluate the members of the expert panel. 

The \textit{weight of the KPI Compliance} $w_{K}$ is defined using:
\begin{equation} \label{equation__weight__5}
\begin{split}
w_{R_{C_{2}}} & = w_{K}(V_{s},T_{s}) \\
& = \frac{1}{N_{K}}\sum_{i=1}^{N_{K}} \frac{K_{C_{i}} \times w_{K_{C_{i}}} }{K_{T_{i}}}
\end{split}
\end{equation}
where $i,N_{K} \in \mathbb{N}$, $w_{K_{C_{i}}}$ represents the confidence weight for the KPI, $K_{C_{i}}$ represents the current KPI calculation, and $K_{T_{i}}$ is the target or estimated KPI for the machine performance during the member working time. The expert panel defines $K_{C_{i}}$ and $K_{T_{i}}$, where $K_{C_{i}}$ is calculated using online machine data, an $K_{T_{i}}$ is set by the team.

The \textit{weight of the user rating} $w_{U}$ is defined using:
\begin{equation} \label{equation__weight__6}
\begin{split}
w_{R_{C_{3}}} & = w_{U} \\
              & = U_{S}
\end{split}
\end{equation}
where $U_{S}$ is the user satisfaction in the range $[0,1]$.

The prior weights for the knowledge rules are composed solely of the expert panel weight, thus:
\begin{equation} \label{equation__weight__7}
\begin{split}
w_{R_{j}} =  w_{P}(V_{t},T_{t}) 
\end{split}
\end{equation}

\begin{figure*}[!htpb]
	\centering
	\includegraphics[width=0.9\textwidth,keepaspectratio]{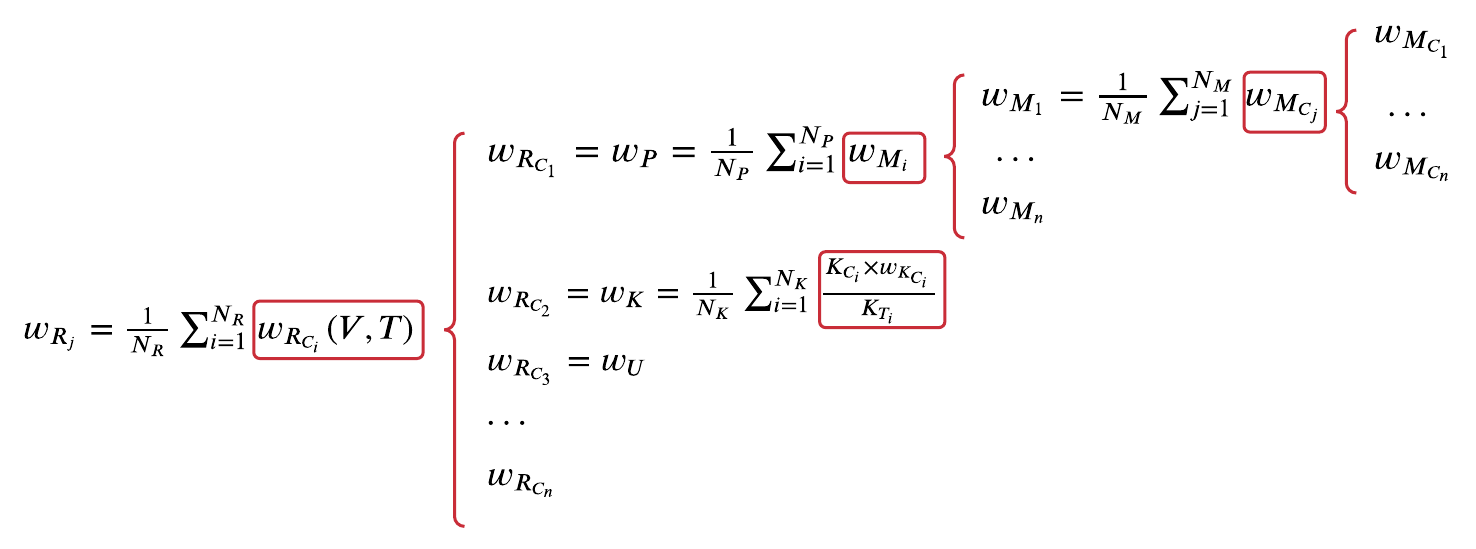}
	\caption{Overview of Confidence Weights}\label{figure__weights_diagram__1}
\end{figure*}

Fig. \ref{figure__weights_diagram__1} shows an overview diagram of the confidence weights. The weight for each rule will have a value in the range $[0,1]$. Though the confidence weight can provide the certainty of the active rule, there is no assessment of the \textit{overall uncertainty}, particularly for knowledge rules that are under the evaluation of acceptance. The Dempster Shafer Evidence Theory (DSET) can support the modeling of \textit{overall uncertainty}.

The knowledge rule, when triggered, will have a value defined in:
\begin{equation} \label{equation__DSET_10}
\begin{split}
\Theta = \{ R_{1},...,R_{n} \}
\end{split}
\end{equation}

Having a triggered rule $R_{i}$:
\begin{equation} \label{equation__DSET_11}
\begin{split}{
	R_{i} = True
	}
\end{split}	
\end{equation}
where $R_{i}$ is the $i_{th}$ focal element of $\Theta$, and $i \in \mathbb{N}$. 

Since the system only triggers one knowledge rule at the time, the BPA weighted sum $S_{wbpa}$ is represented as: 
\begin{equation} \label{equation__DSET_12}
\begin{split}
S_{wbpa} & = m_{R_{j}}*w_{m_{R_{j}}} + U = 1\\
\end{split}
\end{equation}
where $\forall R_{j} \neq True $, $m_{R_{j}}=0$. 
The $S_{wbpa}$ does not take in consideration neither the other focal elements (e.g., knowledge rules that are not active) nor the associated weights. The consideration of the weights of each focal element can improve the quantification of the overall uncertainty. To this end, we present an approximation for the $S_{wbpa}$ called $S_{awbpa}$ that considers all the focal elements using a sensitivity to zero approach.

Cheng et al. \cite{Cheng1988} proposed the use of \textit{sensitivity to zero} when building evidence, which approximates the zero and one values to nearly-zero and nearly-one, respectively. 

\begin{remark}\label{proof__keene__1}
The approximation factor $k$ enhances the evidence definition because all the focal elements are considered, even if these values are nearly zero \cite{Arevalo2017}: 
\begin{equation} \label{equation__DSET_9}
\begin{split}
k = 1-10^{-F}.\\
\end{split}
\end{equation}
where $k \in \mathbb{R}$, and $F \in \mathbb{N}$.
\end{remark}

\begin{proposition}\label{proof__keene__2}
Using the approximation factor $k$ from (\ref{equation__DSET_9}), the BPA weighted sum from (\ref{equation__DSET_4}) is transformed into:
\begin{equation} \label{equation__DSET_13}
\begin{split}
S_{awbpa} & = \sum_{j=1}^{n} m'_{R_{j}}*w_{m_{R_{j}}} + U = 1\\
\end{split}
\end{equation}
where $m'_{R_{j}}$ is BPA using the approximation factor $k$, and is represented as: 
\begin{equation} \label{equation__DSET_14}
     m'_{R_{j}}=\begin{cases}
         k ,\quad\quad \text{if}\ R_{j} = \text{True} \\
         \frac{1-k}{n-1},\quad \text{otherwise}  \\
         \end{cases}
\end{equation}
\end{proposition}

\begin{assumption}\label{proof__keene__3}
Considering a factor F such as: 
$F \gg 1$ , and therefore $k \rightarrow 1$.

\end{assumption}

Similar to Proposition \ref{definition__dempster__2}, the integrity of Equation (\ref{equation__DSET_4}) must be preserved when applying Proposition \ref{proof__keene__2}. Therefore,  Proposition \ref{proof__keene__2} must be consistent with Equation (\ref{equation__DSET_4}) by mathematical proof.  

\begin{lemma}\label{proof__keene__4}
Denote the $S_{bpa}$ and $S_{awbpa}$ as the BPA sum and approximated BPA weighted sum with an explicit overall uncertainty definition, respectively. Then, it holds:
\begin{equation} \label{equation__DSET_15}
\begin{split}
S_{bpa} = S_{awbpa}\\
\end{split}
\end{equation}
\end{lemma}

\renewcommand\qedsymbol{$\blacksquare$}

\begin{proof}
Assuming a factor $F \gg 1$, thus, the approximation factor $k \rightarrow 1$, and therefore, the BPA of the active rule $R_{i}$ will tend $m'_{R_{j}} \rightarrow 1$; whereas, the BPA of the inactive rules will tend to zero. As a result, $R_{i}*w_{R_{i}} + U = R_{i}*w_{R_{i}} + 1 - R_{i}*w_{R_{i}}$, which equals to one, satisfying, thus, Equation (\ref{equation__DSET_3}). 
\end{proof}

The BPA $m'_{R_{j}}$ can be transformed into an array form for posterior calculations \cite{Arevalo2018}: 
\begin{equation} \label{equation__DSET_16}
\begin{split}{
	m_{R_{j}} = [\ 
	 R_{1}*w_{R_{1}}\
	...\ 
	R_{n}*w_{R_{n}}\
	 U\ ] }
\end{split}	
\end{equation}

where $R_{j}$ and $w_{R_{j}}$ are the $j_{th}$ element of $\Theta$ for the rule and confidence weight, respectively; $U$ is the overall uncertainty, and $j,n \in \mathbb{N}$.

\subsection{Knowledge Modeling and Validation}

\subsubsection{Knowledge Modeling}
Now that we have extracted the knowledge using the causal method FMEA, this knowledge can be used as a knowledge model by formalizing the rules as primitive recursive functions.
Kleene \cite{Kleene1971} defined that "a function $\varphi$ is \textit{primitive recursive in} $\psi_{1}$,...,$\psi_{l}$ (briefly $\Psi$), it there is a finite sequence $\varphi_{1}$,...,$\varphi_{k}$ of (occurrences) functions (called a \textit{primitive recursive derivation of $\varphi$ from} $\Psi$) such that each function of the sequence is either one of the functions $\Psi$ (the \textit{assumed functions}), or an initial function, or an immediate dependent or preceding functions, and the last function $\varphi_{k}$ is $\varphi$".

\begin{definition}[Kleene \cite{Kleene1971}]
Kleene defined the switch case function as "a set of predicates $Q_{1}$,...,$Q_{m}$ is \textit{mutually exclusive}, if for each set of arguments not more than one of them is true.
\#F. \textit{The function $\varphi$ defined thus}
\begin{equation} \label{equation__modeling_1}
     \varphi(x_{1},...,x_{n})=\begin{cases}
         \varphi_{1}(x_{1},...,x_{n}) \quad\quad \text{if} \quad Q_{1}(x_{1},...,x_{n}),  \\
         ...\\
         \varphi_{m}(x_{1},...,x_{n}) \quad\quad \text{if} \quad Q_{m}(x_{1},...,x_{n}),  \\
         \varphi_{m+1}(x_{1},...,x_{n}) \quad \text{otherwise} \\
         \end{cases}
\end{equation}
\textit{where $Q_{1}$,...,$Q_{m}$ are mutually exclusive predicates (or $\varphi(x_{1},...,x_{n})$ shall have the value given by the first clause which applies) is primitive recursive in $\varphi_{1}$,...,$\varphi_{m+1}$, $Q_{1}$,...,$Q_{m}$}." 
\end{definition}

The knowledge rules from the extended FMEA were defined as functions of process variables and process thresholds:
\begin{equation} \label{equation__modeling_1a}
    \begin{split}
    R_{j} = f(V_{1},...,V_{n}, T_{1},...,T_{n})
    \end{split}
\end{equation}
where $V_{1},...,V_{n}$ represent the process variables, $T_{1},...,T_{n}$ are the variable thresholds used in the knowledge rules, and $j,n \in \mathbb{N}$. The rules extracted by the extended FMEA are mutually exclusive. The mutual exclusivity property of the rules satisfy the condition of the switch case function of Kleene.

Thus, the knowledge rules can be represented with the function $L_{R}$ (to simplify the equations the term $(V_{1},...,V_{n}, T_{1},...,T_{n})$ will not be written) :
\begin{equation} \label{equation__modeling_2}
    \begin{split}
        L_{R}=\begin{cases}
         L_{R_{1}} \quad\quad \text{if} \quad R_{1},  \\
         ...\\
         L_{R_{m}} \quad\quad \text{if} \quad R_{m},  \\
         L_{R_{m+1}} \quad \text{otherwise},  \\
         \end{cases}
    \end{split}
\end{equation}

where $R_{1},...,R_{m}$ are the knowledge rules,  $L_{R_{1}},...,L_{R_{m}}$ represent the labels correspondent for each rule, $L_{R_{m+1}}$ is the exit clause, and $m \in \mathbb{N}$.

The knowledge-based model can also be represented as:
\begin{equation} \label{equation__modeling_3a}
    \begin{split}
        L_{T_{R}}=\begin{cases}
         L_{T_{R_{1}}}  \quad\quad \text{if} \quad R_{1},  \\
         ...\\
         L_{T_{R_{m}}}  \quad\quad \text{if} \quad R_{m},  \\
         L_{T_{R_{m+1}}} \quad \text{otherwise},  \\
         \end{cases}
    \end{split}
\end{equation}

Where the transformed rule $L_{T_{R_{j}}}$ was defined using Equation (\ref{equation__DSET_9}):
\begin{equation*} 
     L_{T_{R_{j}}}=\begin{cases}
         k ,\quad\quad \text{if}\ R_{j} = \text{True} \\
         \frac{1-k}{n-1},\quad \text{otherwise}  \\
         \end{cases}
\end{equation*}




The next step is the integration of the knowledge-based model and the uncertainty of each rule to determine the confidence level of the rules. The previous section defined the uncertainty as a confidence weight for each rule using Equation (\ref{equation__weight__1}):
\begin{equation*} 
\begin{split}
w_{R_{j}} = \frac{1}{N_{R}}\sum_{i=1}^{N_{R}} w_{R_{C_{i}}}(V,T)
\end{split}
\end{equation*}

Having a triggered rule $R_{j}$:
\begin{equation} \label{equation__modeling_5}
    \begin{split}
        L_{R} = L_{T_{R_{j}}}
    \end{split}
\end{equation}
with its corresponding confidence $w_{R_{j}}$, it provides a relevant assessment of the process, however, the overall uncertainty of the body of knowledge remains unknown. Knowing the uncertainty could provide a perspective on the overall confidence.
Therefore, the overall uncertainty $U$ of the knowledge-based model for the current triggered rule $R_{i}$ must be calculated. For this purpose, the rule $R_{i}$ is transformed into a set of evidence $m_{R_{j}}$ using the equation  (\ref{equation__DSET_6}):
\begin{equation*} 
\begin{split}{
	m_{R_{i}} = [\ L_{w_{R_{j}}}\
	 L_{w_{R_{j+1}}}\
	...\ 
	L_{w_{R_{n-1}}}\
	 U\ ] }
\end{split}	
\end{equation*}

Where the term $L_{w_{R_{j}}}$ can be represented as:
\begin{equation} 
\label{equation__modeling_6}
\begin{split}{
     L_{w_{R_{j}}}= L_{T_{R_{j}}} \times w_{R_{j}}
     }
\end{split}	
\end{equation}

Thus, the overall uncertainty $U$ is represented using the equation (\ref{equation__DSET_5}):
\begin{equation*} 
\begin{split}
U = 1- \sum_{j=1}^{n} L_{w_{R_{j}}}\\
\end{split}
\end{equation*}

\subsubsection{Knowledge Validation}
Having a knowledge model containing explicit knowledge in the form of rules, the next step is to define a validation strategy to evaluate their performance. For this purpose, the knowledge rules are validated using the KPI calculation for a period of time. Thus, the validation of knowledge rule $R_{j}$ is represented by:
\begin{equation} \label{equation__validation1}
\begin{split}
K_{V_{R_{j}}} = \frac{1}{N_{V}}\sum_{i=1}^{N_{V}} \frac{ K_{C_{i}} \times w_{K_{C_{i}}} }{K_{T_{i}}}
\end{split}
\end{equation}

, where $i,N_{V} \in \mathbb{N}$. $w_{K_{C_{i}}}$ represents the confidence weight for the KPI, $K_{C_{i}}$ represents the current KPI calculation, and $K_{T_{i}}$ is the target or estimated KPI for the knowledge rule. The next step was to compare the validation results of the $K_{V_{R_{j}}}$ with a threshold to approve the rule, if successful.
In this study, the rules are evaluated on short-term and long-term bases. The short-term basis evaluates the acceptance of a new knowledge rule; whereas, the long-term basis evaluates the long-term effects.   

\subsection{Interactive Assessment System}
This subsection provides considerations from the software engineering side for the deployment of the KLAFATE as a backend, as well as the user interface using the augmented reality device HoloLens. Software development followed an agile methodology, as the issues were defined and grouped in working package \textit{sprints} for a 2-weeks time slot, keeping a backlog for future tasks. 
The first challenge is to define software requirements. For this purpose, the backend and HoloLens are addressed first separately, and second, it is addressed as a system. 
In the first step, the backend must fulfill the following tasks:
\begin{itemize}
\item Communication to the machine and to the HoloLens
\item Data collection of the machine
\item Knowledge Extraction through FMEA parsing  
\item Uncertainty Quantification of the Knowledge Rules
\item Sending assessment messages to the HoloLens
\item Receiving user rating and report  
\item Updating rule weights
\item Calculating the time response of the system and communications
\end{itemize}

The HoloLens needed to fulfill the following tasks:
\begin{itemize}
\item Display the assessment provided by the backend
\item Request a report from the user in case no effective diagnosis is available
\item Request a user rating
\item Provide a voice command to enhance the user experience
\end{itemize}

Having defined the tasks for the backend and HoloLens, it is possible to sketch a sequence diagram that shows the interactions between the backend and HoloLens. 
\begin{figure}[!htpb]
	\centering
	\includegraphics[width=0.3\textwidth,keepaspectratio]{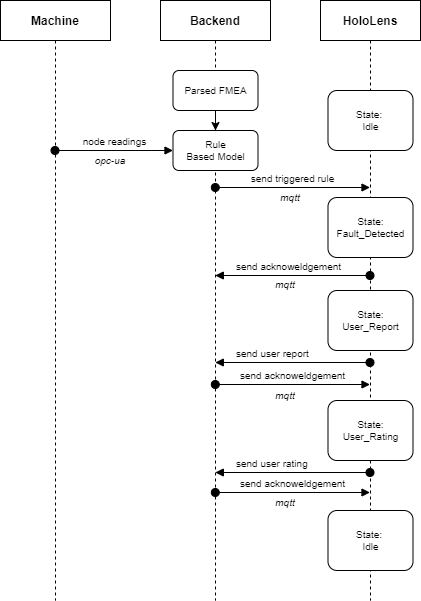}
	\caption{Sequence Diagram of the Interactive Assessment System}\label{figure__proposal_system__1}
\end{figure}

As shown in Figure \ref{figure__proposal_system__1}, the major actors are the machine, backend, and HoloLens. However, the backend has main modules for communication using OPC-UA and MQTT for reading and writing, an MQTT broker, a parsing module to extract information from the FMEA, build the knowledge model, and the main function.
The next step is to define the flow diagrams and pseudo-codes to identify the modules and functions. We considered the use of Git, a version control system, to work collaboratively and keep track of software changes.
Finally, we addressed the hardware, for which we considered a laptop as a device for software development and testing, and HoloLens as the user interface. Having tested the functionality of the backend, it is possible to use different hardware setups to host the backend, such as a cloud platform, locally on a server, or even as edge computing (e.g., using an industrial PC on the shop floor).

\section{Use Case: Application of the KLAFATE in a Laboratory Bulk Good System}\label{section_usecase}

\begin{figure*}[!htbp]
	\centering
	\includegraphics[width=0.65\textwidth,keepaspectratio]{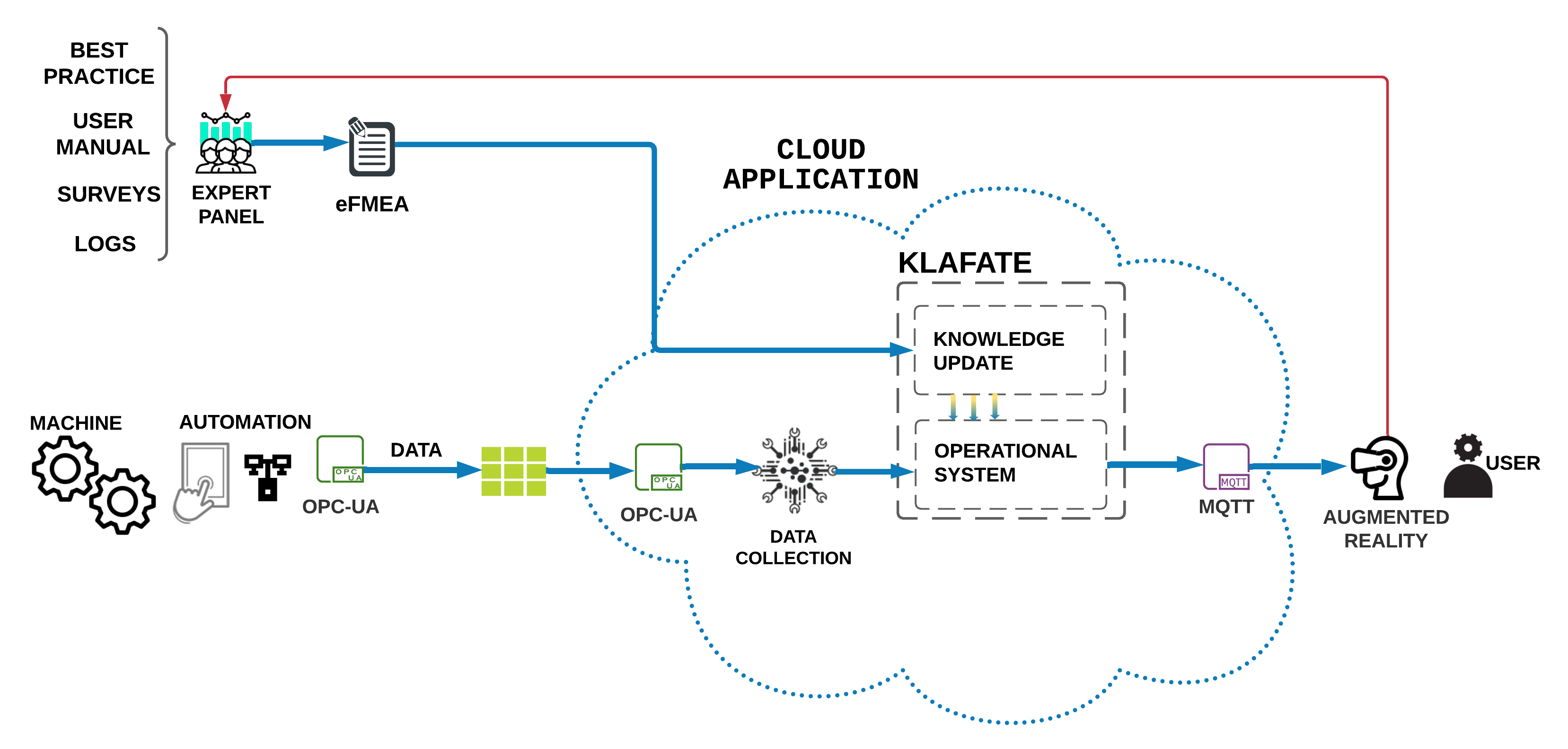}
	\caption{Overview Diagram for the Use Case}\label{figure__use_case_overview_diagram}
\end{figure*}

This section describes the practical implementation of KLAFATE, as well as the test results using a laboratory testbed consisting of a small-scale bulk good system (BGS). This section is divided into the following subsections: BGS description, implementation of KLAFATE, results, and discussion. An overview of this use case is shown in Figure \ref{figure__use_case_overview_diagram}.
The backend hosts the KLAFATE and provides the communication interfaces OPC-UA and MQTT, which are used for the communication of the BGS and augmented reality device \textit{HoloLens}, respectively.

The backend was developed using Python 3.8 in the IDE Anaconda. The backend was tested on a laptop with an Intel Core i7, 32GB RAM, and 475GB HDD. The backend was run in Windows 10 64-Bit.
The interactive user assessment system consists of a backend and augmented reality (AR) device as the user interface. We chose Microsoft's HoloLens 2 as the AR device for user experience (e.g., holographic support, voice command, head/eye/hand tracking, and customized programming for MQTT communication). The HoloLens specifications include a Qualcomm Snapdragon 850 processor with 64-GB of storage and 4-GB DRAM memory running the Windows holographic operating system. The software Unity 3D was used to develop the AR application. This is a cross-platform game engine developed by Unity Technologies. Unity uses the C\# programming language for software development.

 \subsection{Bulk Good System Laboratory Plant}
\begin{figure}[!htpb]
	\centering
	\includegraphics[width=0.25\textwidth,keepaspectratio]{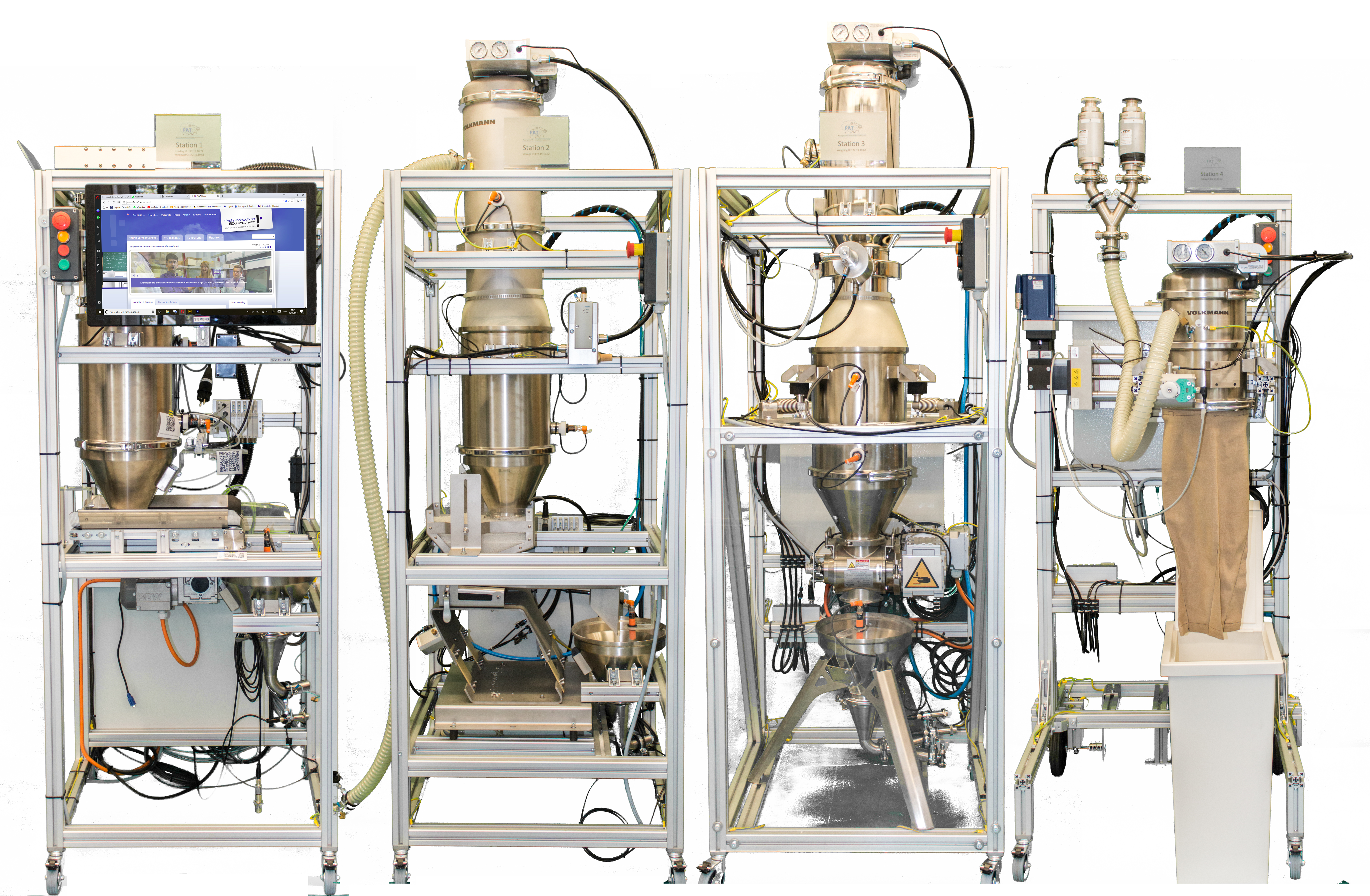}
	\caption{BGS. From left to right: loading, storage, weighing, and filling}\label{figure__use_case_bgs}
\end{figure}

The BGS is a discrete process comprising four stations: loading, storage, weighing, and filling. The BGS uses plastic pellets as bulk goods and possesses common components of a large-scale industrial BGS (e.g., conveyors (motor, vacuum), silos, valves, weighing and dosing stations, and an automation platform). Figure \ref{figure__use_case_bgs} shows the stations in the BGS.

Each station can function standalone or together as a system. The station loading has an industrial PC (IPC) containing a human-machine interface (HMI). The remaining stations possessed a PLC ET200. Each station has an embedded OPC-UA server that facilitates data exchange between the stations and the KLAFATE hosted in the backend.
Table \ref{table__BGS1__use_case} summarizes the list of setpoints for each station. Table \ref{table__BGS2__use_case} lists the variables used in data collection.

\begin{table}[!htbp]
	\centering
	\caption{BGS Operation Parameters of the stations Loading, Storage, Weighing, and Filling.}
	\begin{tabular}{l c c c c}
		\toprule
		\textbf{Variable} &  \textbf{Loading} &  \textbf{Storage} &  \textbf{Weighing} &  \textbf{Filling}\\ 		
		\midrule
		Vacuum Time                   & x & x & x & x\\ 
		Motor RPM                     & x &  &  & \\
		Dosing Motor Weight           &  &  & x & \\
		\bottomrule		 
	\end{tabular}
	\label{table__BGS1__use_case} 
\end{table}

\begin{table}[!htbp]
	\centering
	\caption{BGS Variables of the stations Loading, Storage, Weighing and Filling for data collection.}
	\begin{tabular}{l c c c c}
		\toprule
		\textbf{Variable} &  \textbf{Load.} &  \textbf{Stor.} &  \textbf{Weigh.} &  \textbf{Fill.}\\ 		
		\midrule
		Vacuum Time                   & x & x & x & x\\ 
		Discharge Flap Open Time      & x & x & x & x\\
		Actual Pressure               & x & x & x & x\\
		Filling Height Min State      & x & x & x & \\
		Motor On/Off                  & x &  &  & \\
		Belt Conveyor Actual Speed    & x &  &  & \\
		Belt Conveyor Setpoint Speed  & x &  &  & \\
		Filling Height Max Value      & x & x & x & \\
		Filling Height Min Value      & x & x & x & \\
		Overflow Value                & x & x & x & \\
		Vibration Conveyor            &   & x &  & \\ 
        Out of Weighing Range         &   &  & x & \\
        Dosing Motor Register         &   &  & x & \\        
        Dosing Motor Actual Speed     &   &  & x & \\
        Dosing Motor Setpoint Speed   &   &  & x & \\
        System Mode                   &   &  & x & \\
        Container Available           &   &  & & x \\
		\bottomrule		 
	\end{tabular}
	\label{table__BGS2__use_case} 
\end{table}

\subsection{Experiment Design}
The experiment illustrates the application of KLAFATE in a small-scale industrial testbed. To pursue the experiments, it is necessary to perform a setup procedure in the BGS, KLAFATE, and expert panel.  

The \textit{expert panel setup} included two experts, each with at least 1-year of experience working with the Bulk Good System, and one apprentice with no experience in the machine. The panel discusses and proposes a troubleshooting program and new recipes to optimize the process. These recipes are a collection of machine parameters that allows the machine to achieve the best KPIs. The years of experience of the experienced worker have been exaggerated for illustration purposes in the experiments (see Table \ref{table__use_case__setup__1}). 

\begin{table}[!htbp]
	\centering
	\caption{Expert Panel from the Template \textit{Profile}}
	\begin{tabular}{l c c c c}
		\toprule
		\textbf{Member} & $E_{G}$ &  $E_{M}$ & $w$ & $p$\\ 		
		\midrule
		Operator 1 & 10 & 10 & 0.08 & 2.4 \\ 
		Operator 2 & 5 & 5 & 0.1 & 1.9 \\ 
		Operator 3 & 1 & 0 & 0.25 & 1.4  \\ 
		\bottomrule
	\end{tabular}
	\label{table__use_case__setup__1} 
\end{table}

The \textit{BGS setup} consists of the initial conditions for the stations, such as
the machine parameters, product weight, and compressed air pressure. In addition, before every experiment (e.g., testing a process recipe), the silos were filled to 90\% of its capacity. 


The \textit{KLAFATE setup} defines the constants in the \textit{settings} template. These constants are the thresholds for the rules and confidence weights. Thresholds are grouped into team, system, and component. 
Some of the thresholds for the team, system, and component are listed in Tables \ref{table__use_case__setup__4}, \ref{table__use_case__setup__5} and \ref{table__use_case__setup__6}, respectively. According to the experiment, the calculation time for $KPI_{Compliance}$ was set to 10min, 20min, and 30min .

\begin{table}[!htbp]
	\centering
	\caption{Team Thresholds from the Template \textit{Settings}}
	\begin{tabular}{l c c}
		\toprule
		\textbf{Threshold} &  \textbf{Value} & \textbf{Unit}\\ 		
		\midrule
		$YEARS\_OF\_EXPERIENCE\_MIN$ & 2 & year\\ 
		$YEARS\_OF\_EXPERIENCE\_MAX$ & 5 & year\\ 
		$YEARS\_OF\_EXPERIENCE\_MAC\_MIN$ & 2 & year\\ 
		$YEARS\_OF\_EXPERIENCE\_MAC\_MAX$ & 5 & year\\ 
		$WEIGHT\_YOE\_HIGH$ & 0.95 & -\\ 
		$WEIGHT\_YOE\_MEDIUM$ & 0.75 & -\\ 
		$WEIGHT\_YOE\_LOW$ & 0.75 & -\\ 
		$WEIGHT\_YOTM\_HIGH$ & 0.95 & -\\ 
		$WEIGHT\_YOTM\_MEDIUM$ & 0.75 & -\\ 
		$WEIGHT\_YOTM\_LOW$ & 0.75 & -\\ 
		$KPI\_HIGH$ & 0.95 & -\\
		$KPI\_MEDIUM$ & 0.75 & -\\
		$KPI\_LOW$ & 0.5 & -\\
		$MIN\_WASTE$ & 0.05 & -\\
		$MAX\_WASTE$ & 0.1 & -\\
		$EXCES\_WASTE$ & 0.2 & -\\
		$MIN\_PROD$ & 1.5 & -\\
		$MAX\_PROD$ & 2.7 & -\\
		\bottomrule		 
	\end{tabular}
	\label{table__use_case__setup__4} 
\end{table}

\begin{table}[!htbp]
	\centering
	\caption{System Thresholds from the Template \textit{Settings}}
	\begin{tabular}{l c c}
		\toprule
		\textbf{Threshold} &  \textbf{Value} &  \textbf{Unit}\\ 		
		\midrule
		$LOWEST\_PRODUCTION\_RATE$ & 1.7 & -\\ 
		$HIGHEST\_PRODUCTION\_RATE$ & 5 & -\\ 
		$LOWEST\_SUCTION\_TIME\_QUALITY$ & 2 & s\\ 
		$LOWEST\_PRESSURE$ & 5 & bar\\ 
		\bottomrule		 
	\end{tabular}
	\label{table__use_case__setup__5} 
\end{table}

\begin{table}[!htbp]
	\centering
	\caption{Component Thresholds from the Template \textit{Settings}}
	\begin{tabular}{l c c}
		\toprule
		\textbf{Threshold} &  \textbf{Value} & \textbf{Unit}\\ 		
		\midrule
		$LOADING\_SUCTION\_TIME\_MIN$ & 3000 & s\\ 
		$LOADING\_SUCTION\_TIME\_MAX$ & 10000 & s\\ 
		$LOADING\_DISCHARGE\_TIME\_MIN$ & 1000 & s\\ 
		$LOADING\_DISCHARGE\_TIME\_MAX$ & 6000 &  s\\ 
		\bottomrule		 
	\end{tabular}
	\label{table__use_case__setup__6} 
\end{table}

\subsection{Implementation of the Proposed Methodology} 
The KLAFATE methodology was applied in two stages: offline and online. \textit{Data collection} supported the offline stage. The main script supported the online stage. Around these two scripts, several scripts provide services such as OPC-UA and MQTT communication, fusion functions, parsing functions for the FMEA templates, and auxiliary functions. In addition, an MQTT broker allows communication between the backend and HoloLens.
The offline stage uses the backend data collection module, in which the console is used as the user interface.
The collected data are used 
for the uncertainty quantification of a failure mode and to validate process knowledge (e.g., process recipes).  
The online stage or interactive assessment system consists of the backend and augmented reality application in the HoloLens.
To operate the KLAFATE, the expert panel needs to complete the following steps: filling up the FMEA templates, initializing the BGS, and running the backend and HoloLens application.
The backend contains the MQTT broker for communication to the main script and HoloLens and the main script. The pseudo-code of the main script is displayed in Algorithm \ref{algorithm_use_case__2}. 

\begin{algorithm}
\caption{Backend Algorithm}\label{algorithm_use_case__2}
\begin{algorithmic}[1]
\Procedure{Backend Main Script}{}
\State initialize MQTT communication
\State loading FMEA dictionaries
\State $FIRST\_RUN\gets True$
\State $user\_stop\gets False$
\While{$user\_stop$}
    \If{$FIRST\_RUN$}
        \State load prior weights $w_{R} \gets w_{R_{pr}}$ by Eq.(\ref{equation__weight__7})
    \Else
        \State read OPC-UA variables from BGS
        \State evaluating active system FM $FM_{S}$
        \If{$FM_{S}$}
            \State $normal \gets False$
        \Else
            \State $normal \gets True$
        \EndIf
        \State evaluating component FM $FM_{C}$
        \State $N_{FM_{C}} \gets $ count of active $FM_{C}$ 
        \For{$i=0$ to $N_{FM_{C}}$}
            \If{$ FM_{C} = FM_{S}$}
                \State $List_{FM_{C}} \gets FM_{C}$ 
            \EndIf  
        \EndFor
        \State transform evidence $E_{FM_{S}}$ by Eq. (\ref{equation__DSET_9})-(\ref{equation__DSET_16})
        \State calculating overall uncertainty $U_{E_{FM_{S}}}$
        \For{$i=0$ to $N_{FM_{C}}$}
                \State retrieving causes $C_{FM_{C}}$ for $FM_{C}$
                \State retrieving recommend. $R_{FM_{C}}$ for $FM_{C}$
                \State $message \gets C_{FM_{C}},R_{FM_{C}}$
        \EndFor
        \State $message \gets FM_{S},w_{R},E_{FM_{S}},U_{E_{FM_{S}}} $  
        \State sending $message$ using MQTT to HoloLens
        \State waiting for ACK from HoloLens $A_{Ho}$
        \State waiting for end from HoloLens $End_{Ho}$
    \EndIf
    \If{$A_{Ho}$ AND NOT $normal$ AND $End_{Ho}$} 
        \State waiting for user report
        \State storing user report
    \EndIf
    \If{$A_{Ho}$ AND $normal$}
        \State sending solved to HoloLens
        \State waiting for $Solved_{Ho}$ from HoloLens
        \If{$Solved_{Ho}$}
            \State update weight $w_{R}$ of the FM by Eq. (\ref{equation__weight__1}) 
        \EndIf
    \EndIf
\EndWhile\label{backendwhile}
\State \textbf{return} 
\EndProcedure
\end{algorithmic}
\end{algorithm}

The HoloLens application contains an interactive assessment application that provides information regarding triggered failure modes. The HoloLens application provides different (internal) services, such as voice commands, recognition of hand gesture, and customized programmed services (e.g., MQTT client, state machines, and handshake communication with the backend).
The system latency was also calculated using the backend. 


\subsection{Results}
This section shows the functionality of KLAFATE: a data collection script for data analysis (e.g., 
validation of new recipes and uncertainty quantification),
data storage of the system time response, and an interactive assessment system through the HoloLens and the backend.

\subsubsection{Example using a Failure Mode at the System Level}
This section provides an example of KLAFATE. For this purpose, we chose a failure mode (FM) at the system level, specifically, \textit{low\_quality\_status}. Table \ref{figure__use_case_implementation__1} lists the extended FMEA at the system level for this FM, whereas Table \ref{figure__extended_fmea_use_case_component} lists the extended FMEA at the component level. 

\begin{table*}[!htpb]
	\centering
	\caption{Recipe Validation Experiment}\label{figure__use_case_implementation__1}
	\includegraphics[width=0.95\textwidth,keepaspectratio]{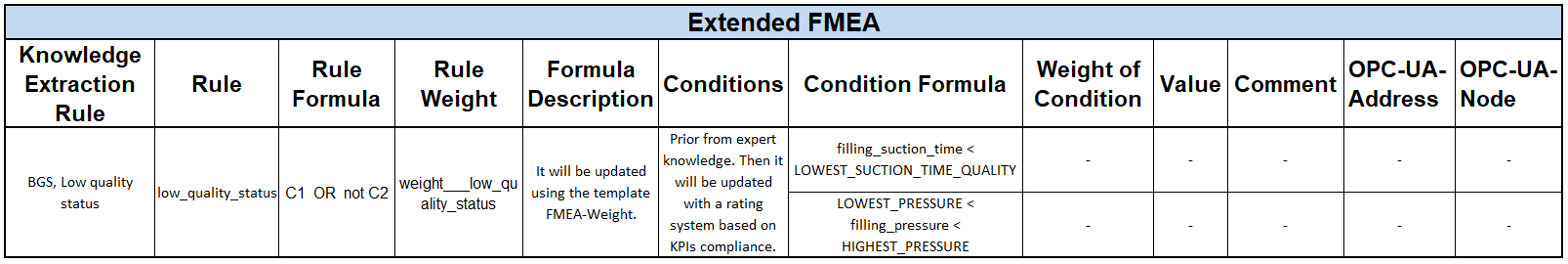}
\end{table*}

\begin{table*}[!htpb]
	\centering
	\caption{Extended FMEA at the component level}\label{figure__extended_fmea_use_case_component}
	\includegraphics[width=0.97\textwidth,keepaspectratio]{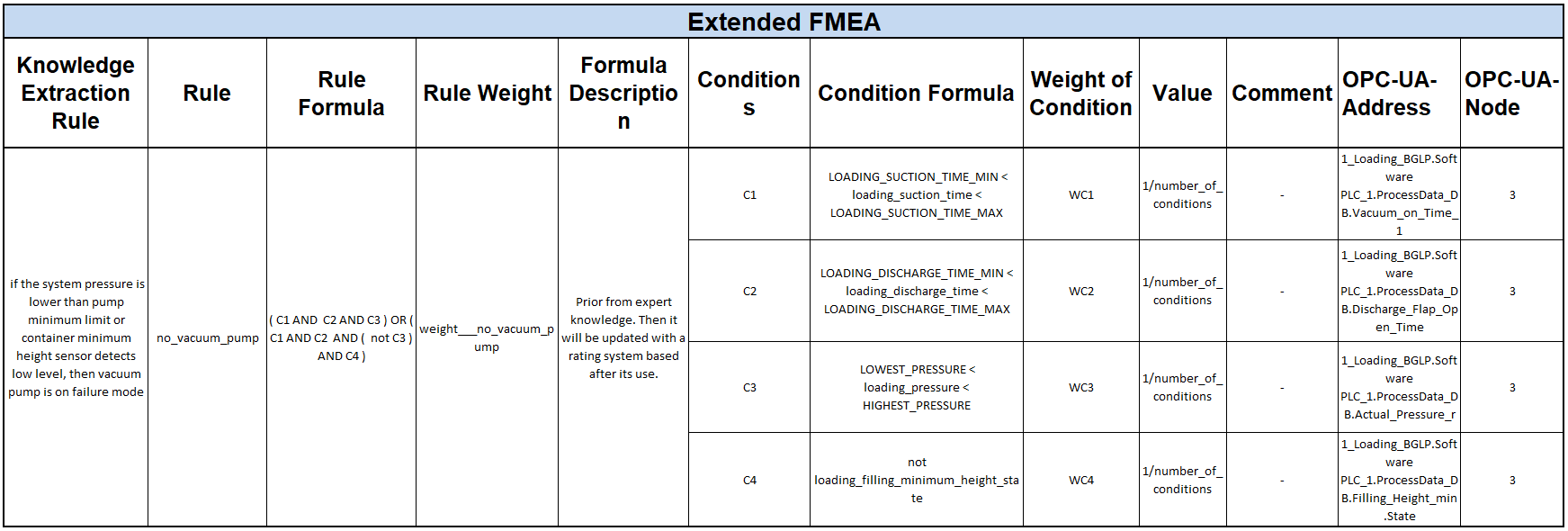}
\end{table*}

\begin{table*}[!htpb]
	\centering
    \caption{Update of Rule Weights}\label{figure__fmea_dynamic_update_weight_use_case}
	\includegraphics[width=0.97\textwidth,keepaspectratio]{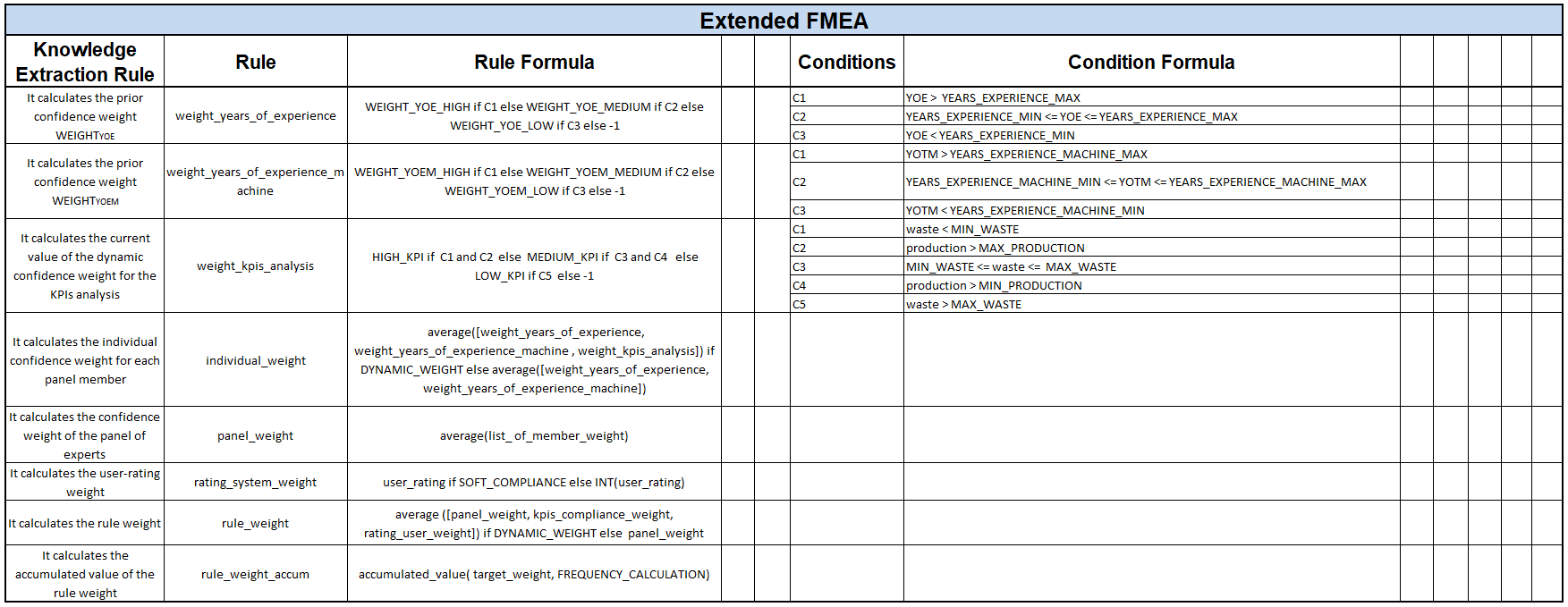}
\end{table*}

The rule \textit{low\_quality\_status} is built using the logic presented in Equations (\ref{equation__model_1}) - (\ref{equation__model_6}).
Defining the active rule \textit{low\_quality\_status} at the system level (see Table \ref{figure__use_case_implementation__1} for the system FMEA) using Equation (\ref{equation__model_1}):
\begin{equation*} \label{equation__example__1}
\begin{split}
R_{i} & = (\ P_{j}\ O_{j}\ P_{j+1} )  \\
      & = C_{1}\ or\ not\ C_{2} \\
\end{split}
\end{equation*}
where $C_{1}$ and $C_{2}$ are defined in Table \ref{figure__use_case_implementation__1}, and the causes and recommendations for the active rule \textit{low\_quality\_status} are provided by the active FM at the component level (see Table \ref{figure__extended_fmea_use_case_component}).
Thus, the active FM \textit{no\_vacuum\_pump} at the component level can be defined using Equation (\ref{equation__model_1}):
\begin{equation*} \label{equation__example__2}
\begin{split}
R_{i} & = ( P_{j}\ O_{j}\ P_{j+1} )  \\
      & = \bigg( P_{j}\ O_{j}\ \Big(P_{k}\ O_{k}\ (P_{k+1})\ O_{k+1}\ P_{k+2}\ O_{k+2}\ P_{k+3} \Big) \bigg) \\
\end{split}
\end{equation*}
where:
\begin{equation*} \label{equation__example__3}
\begin{split}
P_{j} = ( C1\ and\  C2\ and\ C3\ )  \quad\quad O_{j} = or 
\end{split}
\end{equation*}
\begin{equation*} \label{equation__example__3a}
\begin{split}
P_{j+1} & = \Big( P_{k}\ O_{k}\ (P_{k+1})\ O_{k+1}\ P_{k+2}\ O_{k+2}\ P_{k+3} \Big)  
\end{split}
\end{equation*}

where:
\begin{equation*} \label{equation__example__4}
\begin{split}
P_{k} = C_{1} \quad
O_{k} = and \quad
P_{k+1} = C_{2} \quad 
O_{k+1} = and 
\end{split}
\end{equation*}
\begin{equation*} \label{equation__example__5}
\begin{split}
P_{k+2} = not\ C_{3} \quad 
O_{k+2} = and \quad
P_{k+3} = C_{4} \quad
\end{split}
\end{equation*}

Thus, $R_{i}$ can be represented as:
\begin{equation*} \label{equation__example__6}
\begin{split}
R_{i} & = \bigg( \Big(C1\ and\  C2\ and\ C3\ \Big)\ or\ \\
      &\quad\quad \Big( C1\ and\ C2\  and\ (  not\ C3 )\ and\ C4 \Big) \bigg)   \\
\end{split}
\end{equation*}
Where $C_{1}$ - $C_{4}$ are defined in Table \ref{figure__extended_fmea_use_case_component}.
The weight of the rule \textit{low\_quality\_status} can be defined using Equations (\ref{equation__weight__1}), (\ref{equation__weight__3})-(\ref{equation__weight__5}):
\begin{equation*} 
\begin{split}
w_{R_{j}} & = \frac{1}{N_{R}}\sum_{i=1}^{N_{R}} w_{R_{C_{i}}}(V,T) = \frac{1}{3} (w_{P} + w_{K} +  w_{U})  \\
\end{split}
\end{equation*}

\begin{figure*}[!htpb]
	\centering
	\includegraphics[width=0.8\textwidth,keepaspectratio]{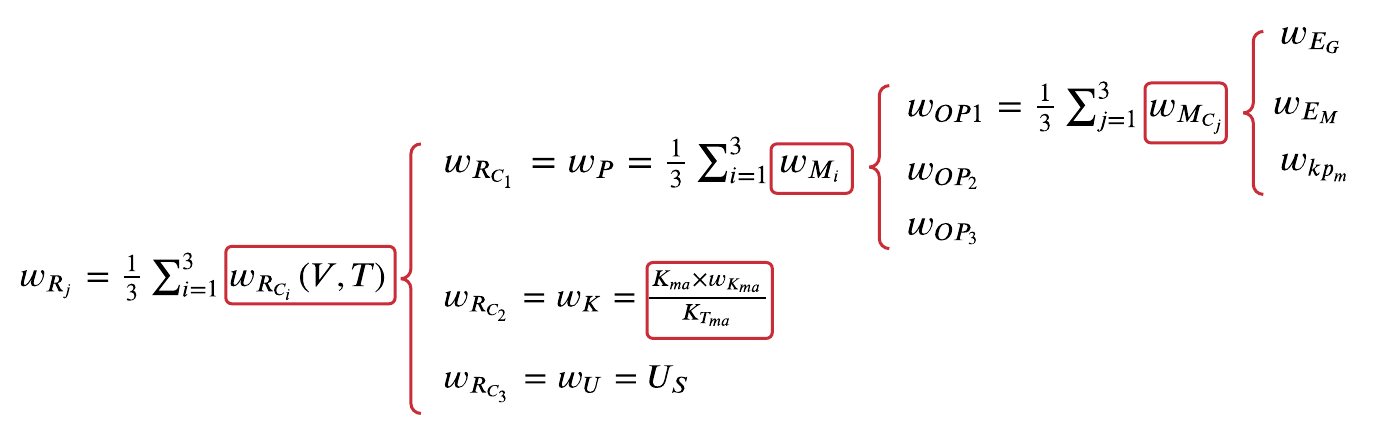}
	\caption{Overview of Confidence Weights}\label{figure__weights_diagram__use_case__1}
\end{figure*}

Likewise, the weight of the rule, it can be modeled using the previous procedure. Fig. \ref{figure__weights_diagram__use_case__1} shows the overview diagram of the confidence weights for the use case. 

Table \ref{table__use_case__setup__1} shows the team setup, where three operators $op_{1}$, $op_{2}$, and $op_{3}$ constitute the expert panel ($N_{P}=3$). 
The weight panel $w_{P}$ can be calculated using the equations (\ref{equation__weight__3}), (\ref{equation__weight__4}):
\begin{equation*} 
\begin{split}
w_{R_{C_{1}}} & = w_{P}(V_{t},T_{t}) = \frac{1}{N_{P}}\sum_{i=1}^{N_{P}} w_{M_{i}} \\
              & = \frac{1}{3} (w_{M_{op_{1}}} + w_{M_{op_{2}}} + w_{M_{op_{3}}}) 
\end{split}
\end{equation*}

Each operator weight $w_{M_{i}}$ is calculated using the criteria $w_{M_{C_{j}}}$: years of experience in general $E_{G}$, years of experience in machine $E_{M}$, and individual performance $K_{A}$ calculated from the KPIs \textit{waste} $w$ and \textit{production rate} $p$ (see Table \ref{table__use_case__setup__1}). 
Where each operator weight $w_{M_{i}}$ is defined using Equation (\ref{equation__weight__4}):
\begin{equation*} 
\begin{split}
w_{M_{i}} & = \frac{1}{N_{M}}\sum_{j=1}^{N_{M}} w_{M_{C_{j}}} = \frac{1}{3} (w_{E_{G}} + w_{E_{M}} + w_{kp_{m}})\\
\end{split}
\end{equation*}

The formulas for $w_{E_{G}}$, $w_{E_{M}}$, and $w_{K_{A}}$ are defined in Table \ref{figure__fmea_dynamic_update_weight_use_case}.
Thus, the weight for the years of experience $w_{E_{G}}$ can be represented using:
\begin{equation*} 
\begin{split}
w_{E_{G}} & = WEIGHT\_YOE\_HIGH\ if\ C1\ else \\
          &\quad\quad WEIGHT\_YOE\_MEDIUM\ if\ C2 \\ 
          & else\ WEIGHT\_YOE\_LOW\ if\ C3\ else\ -1
\end{split}
\end{equation*}
where $C_{1}$, $C_{2}$, and $C_{3}$ are described in Table \ref{figure__fmea_dynamic_update_weight_use_case}.

The weight of the years of experience in the machine $w_{E_{M}}$ can be represented using:
\begin{equation*} 
\begin{split}
w_{E_{M}} & = WEIGHT\_YOEM\_HIGH\ if\ C1\ else \\
          &\quad\quad WEIGHT\_YOEM\_MEDIUM\ if\ C2 \\ 
          & else\ WEIGHT\_YOEM\_LOW\ if\ C3\ else\ -1
\end{split}
\end{equation*}
where $C_{1}$, $C_{2}$, and $C_{3}$ are described in Table \ref{figure__fmea_dynamic_update_weight_use_case}.

Finally, the weight for the KPI performance $w_{kp_{m}}$ can be represented using:
\begin{equation*} 
\begin{split}
w_{K_{A}} & =  HIGH\_KPI\ if\  C1\ and\ C2\  else  \\
          & MEDIUM\_KPI\ if\  C3\ and\ C4\ \\
          & else\ LOW\_KPI\ if\ C5\  else\ -1
\end{split}
\end{equation*}
where $C_{1}$, $C_{2}$, $C_{3}$, $C_{4}$, and $C_{5}$ are described in Table \ref{figure__fmea_dynamic_update_weight_use_case}.

The results are summarized in Table \ref{table__use_case__panel_weight}.
\begin{table}[!htbp]
\centering
\caption{Dynamic Confidence Weights for the Expert Panel}
\begin{tabular}{l c c c c}
	\toprule
	\textbf{Member} & $w_{E_{G}}$ &  $w_{E_{M}}$ & $w_{K_{A}}$ & $w_{M}$ \\ 		
	\midrule
	Operator 1 & 0.95 & 0.95 & 0.75 & 0.88 \\ 
	Operator 2 & 0.75 & 0.75 & 0.75 & 0.75 \\ 
	Operator 3 & 0.5 & 0.5 & 0.5 & 0.5 \\ 
	\bottomrule		 
\end{tabular}
\label{table__use_case__panel_weight} 
\end{table} 

Thus, the weight panel $w_{P}$ can be calculated as:
\begin{equation*} 
\begin{split}
w_{P} & = \frac{1}{3} (w_{op_{1}} + w_{op_{2}} +  w_{op_{3}})  = 0.88 + 0.75 + 0.5 \\
      & = 0.71
\end{split}
\end{equation*}

Using the machine availability $K_{ma}$ as KPI ($N_{K}=1$), and assuming that the assessment solved the problem ($K_{ma}=1$, $K_{T_{ma}}=1$, and $w_{K_{ma}}=1$), the KPI compliance can be calculated using (\ref{equation__weight__5}): 
\begin{equation*} 
\begin{split}
w_{R_{C_{2}}} & = w_{K}(V_{s},T_{s}) = \frac{1}{N_{K}}\sum_{i=1}^{N_{K}} \frac{K_{C_{i}} \times w_{K_{C_{i}}} }{K_{T_{i}}} \\
              & = \frac{K_{C_{1}} \times w_{K_{C_{1}}} }{K_{T_{1}}} = \frac{K_{ma} \times w_{K_{ma}} }{K_{T_{ma}}} = 1.0
\end{split}
\end{equation*}
If the problem has no appropriate diagnosis or cannot be solved, KPI is $K_{ma}=0$.

Assuming a satisfied operator ($U_{S}=0.8$), the user rating weight can be calculated using (\ref{equation__weight__6}): 
\begin{equation*} 
\begin{split}
w_{R_{C_{3}}} = w_{U} = U_{S} = 0.8
\end{split}
\end{equation*}

Thus, the weight for the rule $R_{LQ}$ is calculated using:
\begin{equation*} 
\begin{split}
w_{R_{j}} & = \frac{1}{3} (w_{P} + w_{K} +  w_{U}) = \frac{1}{3} (0.71 + 1.0 + 0.8) = 0.84
\end{split}
\end{equation*}

Assuming that the FMEA system has three FMs: \textit{low\_quality\_status} (LQ), \textit{high\_quality\_low\_production\_status} (LP),  and \textit{high\_quality\_normal\_production\_status} (NP), the knowledge-based model can be represented using equation (\ref{equation__modeling_3a}): 
\begin{equation*} 
    \begin{split}
        L_{T_{R}}=\begin{cases}
         L_{T_{R_{LQ}}} \quad\quad \text{if} \quad R_{LQ},  \\
         L_{T_{R_{LP}}} \quad\quad \text{if} \quad R_{LP},  \\
         L_{T_{R_{NP}}} \quad\quad \text{if} \quad R_{NP},  \\
         L_{T_{R_{E}}} \quad \text{otherwise},  \\
         \end{cases}
    \end{split}
\end{equation*}
The weights of the rules $w_{R_{LP}}$ and $w_{R_{NP}}$ will be assumed to be the prior weights using (\ref{equation__weight__7}):
\begin{equation*} 
\begin{split}
w_{R_{LP}} = w_{R_{NP}} =  w_{P}(V_{t},T_{t}) = 0.71
\end{split}
\end{equation*}

The knowledge-based model triggers rule $R_{LQ}$, which can be transformed into $ L_{w_{R}}$ using (\ref{equation__modeling_5}):
\begin{equation*} 
    \begin{split}
        L_{w_{R}} & = L_{T_{R_{LQ}}} \times w_{R_{LQ}}
    \end{split}
\end{equation*}

where $L_{T_{R_{LQ}}}$ is defined using Equation (\ref{equation__DSET_8}):
\begin{equation*} 
     L_{T_{R_{LQ}}}=\begin{cases}
         k ,\quad\quad \text{if}\ R_{LQ} = \text{True} \\
         \frac{1-k}{n-1},\quad \text{otherwise}  \\
         \end{cases}
\end{equation*}
Assuming an $F=2$, the approximation factor $k$ is calculated using equation (\ref{equation__DSET_4}): 
\begin{equation*} 
\begin{split}
    k = 1-10^{-F} = 1-10^{-2} = 0.99
\end{split}
\end{equation*}
Thus, since $R_{LQ}$ is active, $L_{T_{R_{LQ}}}$ yields:
\begin{equation*} 
\begin{split}
     L_{T_{R_{LQ}}} = k = 0.99 
\end{split}
\end{equation*}
The rest of the rules are calculated using:
\begin{equation*} 
\begin{split}
     L_{T_{R_{LP}}} = L_{T_{R_{NP}}} = \frac{1-k}{n-1}  =  \frac{1-0.99}{3-1} = 0.005
\end{split}
\end{equation*}

The uncertainty of the system is calculated using the evidence theory, where the power set is represented using Equation (\ref{equation__DSET_2a}):
\begin{equation*} 
\begin{split}
\Theta = \{LQ, LP, NP\}
\end{split}
\end{equation*}

The rule $L_{w_{R}}$ can be transformed into a set of evidence $m_{R_{LQ}}$ using the equation (\ref{equation__DSET_6}):
\begin{equation*} 
\begin{split}
	m_{L_{w_{R}}} & = [\ L_{w_{R_{LQ}}}\quad L_{w_{R_{LP}}}\quad L_{w_{R_{NP}}}\quad U\ ] 
\end{split}	
\end{equation*}

which can also be represented as:
\begin{equation*} 
\begin{split}
	m_{R_{LQ}} & = [L_{T_{R_{LQ}}} \times w_{R_{LQ}}\quad  \ L_{T_{R_{LP}}} \times w_{R_{LP}}\quad \\
               & \quad\quad L_{T_{R_{NP}}} \times w_{R_{NP}}\quad U\ ] \\
\end{split}	
\end{equation*}

where the overall uncertainty $U$ is represented using the equation (\ref{equation__DSET_8}):
\begin{equation*} 
\begin{split}
U & = 1 - \sum_{j=1}^{n} L_{w_{R_{j}}} \\
  & = 1 - \sum_{j=1}^{n} L_{T_{R_{j}}} \times w_{R_{j}}\\
  & = 1 - (L_{T_{R_{LQ}}} \times w_{R_{LQ}} + \ L_{T_{R_{LP}}} \times w_{R_{LP}}\ \\
  & \quad\quad + L_{T_{R_{NP}}} \times w_{R_{NP}}) \\
  & = 1 - ( 0.99*0.84 + 0.005*0.71 + 0.005*0.71 ) \\
  & = 0.16
\end{split}
\end{equation*}

Thus, the set of evidence is calculated as:
\begin{equation*} 
\begin{split}
	m_{R_{LQ}} & = [L_{T_{R_{LQ}}} \times w_{R_{LQ}}\quad \ L_{T_{R_{LP}}} \times w_{R_{LP}}\quad \\
               & \quad\quad L_{T_{R_{NP}}} \times w_{R_{NP}}\quad U\ ] \\
	           & = [0.99*0.84\quad 0.005*0.71\quad 0.005*0.71\quad 0.16] \\
	           & = [0.83\quad 0.003 \quad 0.003\quad 0.16]
\end{split}	
\end{equation*}

Thus, the active rule $R_{LQ}$ has a confidence level of 83\%, and the overall uncertainty lies by 16\%.

\subsubsection{Uncertainty Representation}
\begin{figure}[!htpb]
	\centering
	\includegraphics[width=0.4\textwidth,keepaspectratio]{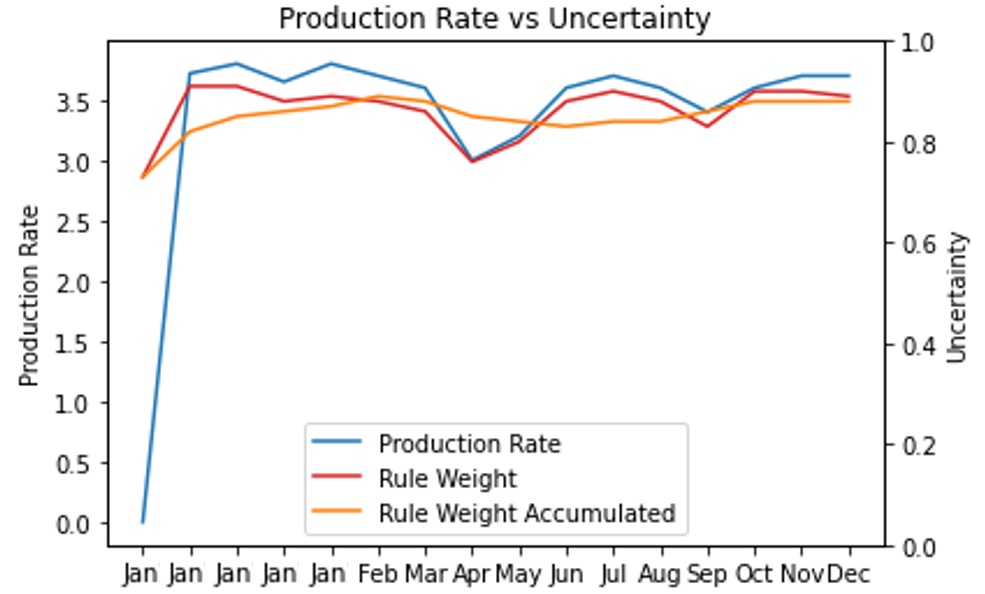}
	\caption{Production rate against uncertainty representation for the rule weight}\label{figure__use_case_uncertainty__1}
\end{figure}

The weight $w_{R}$ represents the uncertainty of the rule $R$ (e.g., a process recipe). The rule was evaluated at intervals of 10min, 20min, and 30min. Assuming a steady process, synthetic data are created to illustrate the change in the weight over time. For this purpose, the panel weight $w_{p}$ was assumed to change twice a year, assuming a regular evaluation (e.g., the operators received training). The production rate was assumed to be steady with an average value of 3.5 prod/min. However, external disturbances (e.g., material shortages or pressure decay) were considered during April and September, as shown in Fig. \ref{figure__use_case_uncertainty__1}. 
These fluctuations in the production rate also influenced the user rating weight $w_{u}$, which is, in this case, the user satisfaction that was not fulfilled (e.g., the estimation of the KPI was not reached). 
The weight $w_{R}$ follows fluctuations in the production rate, as shown in Figure \ref{figure__use_case_uncertainty__1}. In contrast, the accumulated weight $w_{R_{a}}$ has a steadier trend absorbing the disturbances.

\subsubsection{Knowledge Validation}
\begin{figure*}[!htbp]
	\centering
	\begin{subfigure}[b]{0.3\textwidth}
    	\includegraphics[width=\textwidth,keepaspectratio]{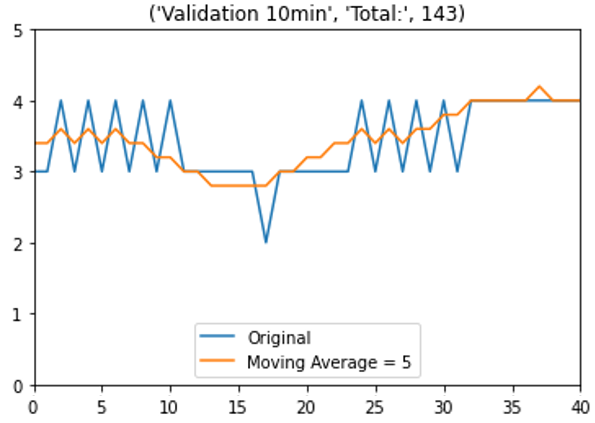}
    	\caption{10 Minutes}\label{figure__use_case_validation__1}
	\end{subfigure}
	\begin{subfigure}[b]{0.3\textwidth}
    	\includegraphics[width=\textwidth,keepaspectratio]{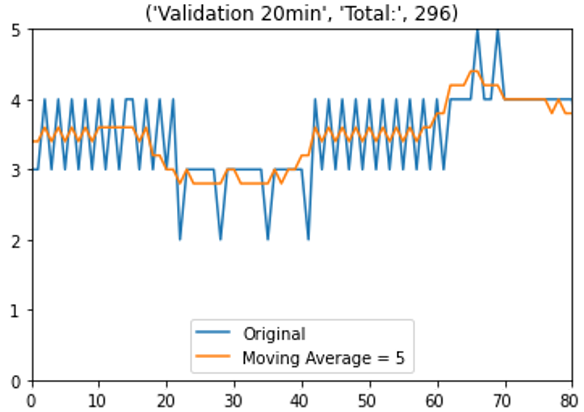}
    	\caption{20 Minutes}\label{figure__use_case_validation__2}
	\end{subfigure}
	\begin{subfigure}[b]{0.3\textwidth}
    	\centering
    	\includegraphics[width=\textwidth,keepaspectratio]{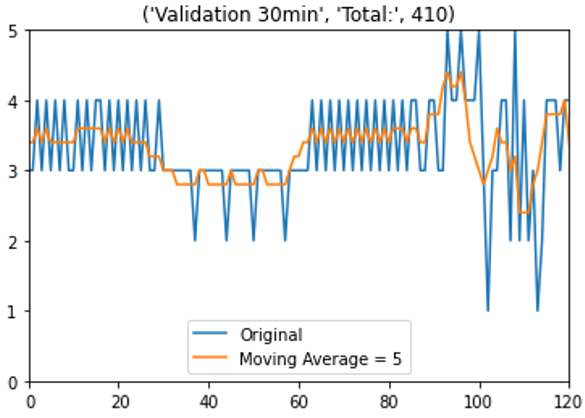}
    	\caption{30 Minutes}\label{figure__use_case_validation__3}	
	\end{subfigure}
	\caption{Recipe Validation Experiment 1}\label{figure__use_case_validation__4}
\end{figure*}

The scenario illustrates the machine operation by an inexperienced operator and using KLAFATE.
The KPI under observation is the production rate \textit{PR} measured in [prod/min]. This experiment was conducted using different time slots: 10, 20, and 30 min for the KPI calculations. The data collection script is used to evaluate the performance of the machine by comparing the current production rate with user estimation.
The experiment began with a steady or normal condition on the machine. This machine condition corresponds to the label \textit{NP} or \textit{normal production}, which corresponds to an average production rate of 3.4 prod/min for the 30 min time slot, as shown in Figure \ref{figure__use_case_validation__1}.
An inexperienced operator sets a new recipe \textit{X1} in the machine with an estimated production rate of 4 prod/min, corresponding to an improvement of 18\%. This recipe yielded a production rate of 2.9 prod/min. The recipe did not reach the estimation or the current normal production rate \textit{NP}; therefore, the recipe was discarded and the operator loaded the previous recipe \textit{NP}. 
The expert panel suggested a new recipe \textit{X2} in the machine with an estimated production rate of 4.2 prod/min, which corresponds to an improvement of 23\%. Observing the plot in Fig. \ref{figure__use_case_validation__1}, using the new recipe fulfills the estimation using a \textit{time slot of 10min with a moving average of five samples}. However, evaluating the time slots for 20 min and 30 min (Figures \ref{figure__use_case_validation__2}, and \ref{figure__use_case_validation__3} respectively), the production rate decays. The reason for this decay relies on the silo levels, which cannot be filled by the selected suction time of 3s. This effect can only be observed over a long evaluation period. Thus, using a time slot of 30 min, the production rate did not fulfill the estimation. However, the new recipe yields a better production rate than the current recipe. Nevertheless, to approve a new recipe, a new analysis should be performed using a longer time slot.

\begin{figure*}[!htbp]
	\centering
	\begin{subfigure}[b]{0.3\textwidth}
    	\includegraphics[width=\textwidth,keepaspectratio]{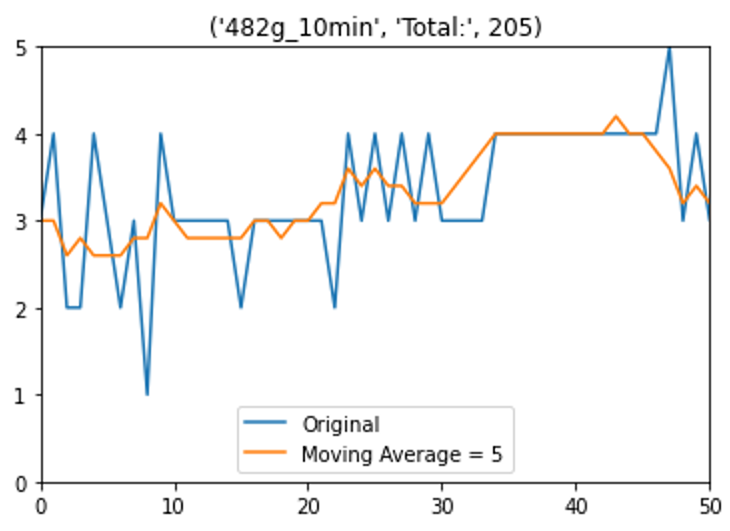}
    	\caption{10 Minutes}\label{figure__use_case_optimization__1}
	\end{subfigure}
	\begin{subfigure}[b]{0.3\textwidth}
    	\includegraphics[width=\textwidth,keepaspectratio]{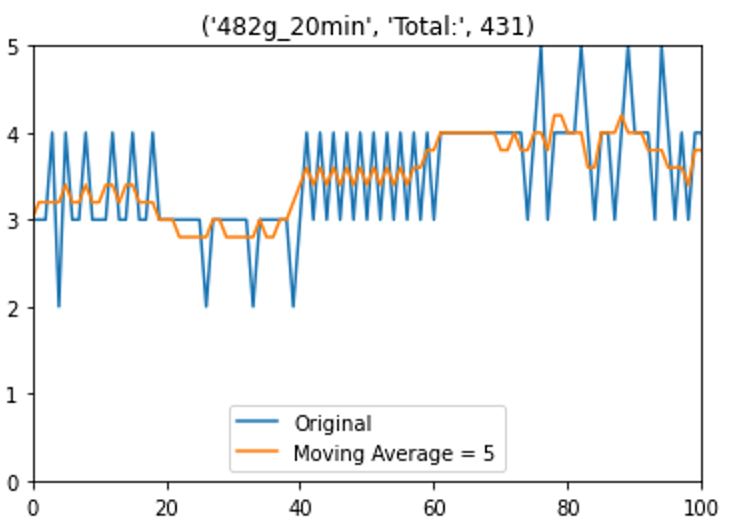}
    	\caption{20 Minutes}\label{figure__use_case_optimization__2}
	\end{subfigure}
	\begin{subfigure}[b]{0.3\textwidth}
    	\centering
    	\includegraphics[width=\textwidth,keepaspectratio]{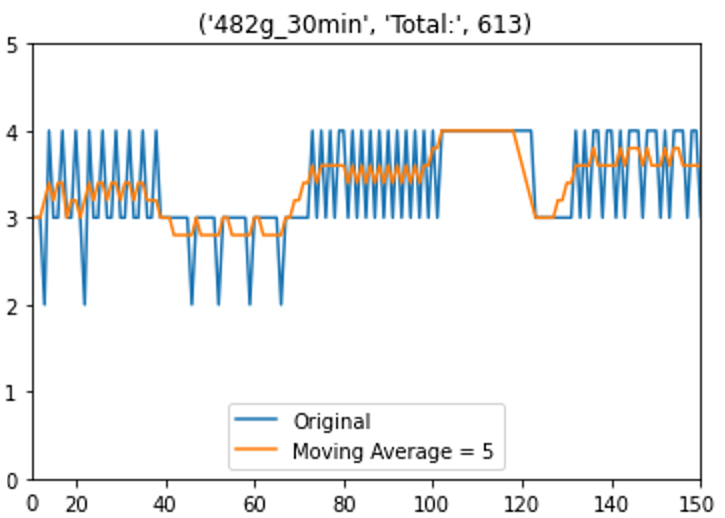}
    	\caption{30 Minutes}\label{figure__use_case_optimization__3}	
	\end{subfigure}
	\caption{Recipe Validation Experiment 2}\label{figure__use_case_validation__5}
\end{figure*}

The second experiment, as seen in Figure \ref{figure__use_case_validation__5}, begins with a low production \textit{LP} condition on the machine. This condition corresponds to an average production rate of 3.2 prod/min. Similarly, an inexperienced operator sets a new recipe \textit{X3} to improve the production rate; however, this also causes the production rate to dip, as seen in the previous experiment. In contrast, this experiment changes the recipe to \textit{NP} after detecting the failure of the recipe \textit{X3}. This recipe resulted in an increase in the production rate by 6\%. The expert panel then suggests a new recipe \textit{X4} in the machine, with an estimated production of 4 prod/min. Similar to the previous experiment, the initial estimate was achieved. However, as it progressed towards 30 minutes, the production rate started to decay. The cause of this decay is similar to what was discussed in the first experiment. As a result, another recipe \textit{X5} was suggested by the expert panel, where it manages to stop the decay and slightly increase the production by 6\% compared to \textit{X3}.

A one-way ANOVA test was conducted to determine whether there were significant differences between the production rates of the recipes. An initial null hypothesis and an alternative hypothesis were set, where the null hypothesis stated that all setpoints yielded the same production rates. By contrast, the alternative hypothesis contradicts this by stating that there is a difference between production rates. This test was done on the three recipes. Given an alpha of 0.05, the resulting p-value was \sn{3.12}{-23}, indicating that the null hypothesis should be rejected. Therefore, it can be concluded that the three recipes did not yield the same production rates and had significant differences.  

\subsubsection{Time Response of the System}
The time response of the system is evaluated using different perspectives: the communication between the backend and HoloLens, internal backend cycle, and time required for failure mode detection and user assessment.
The latency of the MQTT communication had an average of 1s; however, the last trials had an average of less than 1s. This time is required to send the assessment message from the backend to its reception in the HoloLens.
The time required from failure mode detection until visualization on the HoloLens had an average of 5s.
The internal cycle time of the backend had an average of 10s. This cycle time also depends on the user interaction, which implies that the user can influence this time measurement (e.g., an inexperienced user would require additional time to evaluate the recommendation).


\subsubsection{Interactive Assessment System using the backend and the HoloLens}
\begin{figure*}[!htbp]
	\centering
	\begin{subfigure}[b]{0.45\textwidth}
    	\includegraphics[width=\textwidth,keepaspectratio]{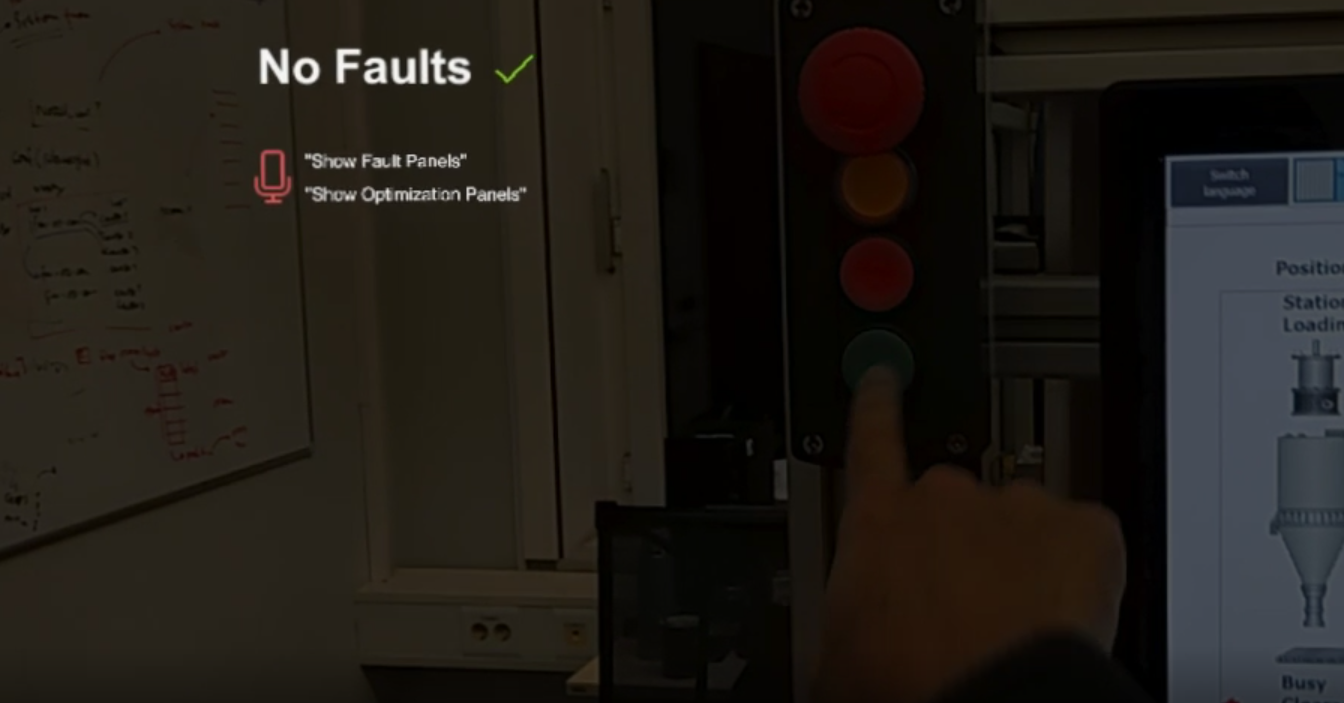}
    	\caption{HoloLens under normal conditions}\label{figure__use_case_hololens__1}
	\end{subfigure}
	~
	\begin{subfigure}[b]{0.45\textwidth}
    	\includegraphics[width=\textwidth,keepaspectratio]{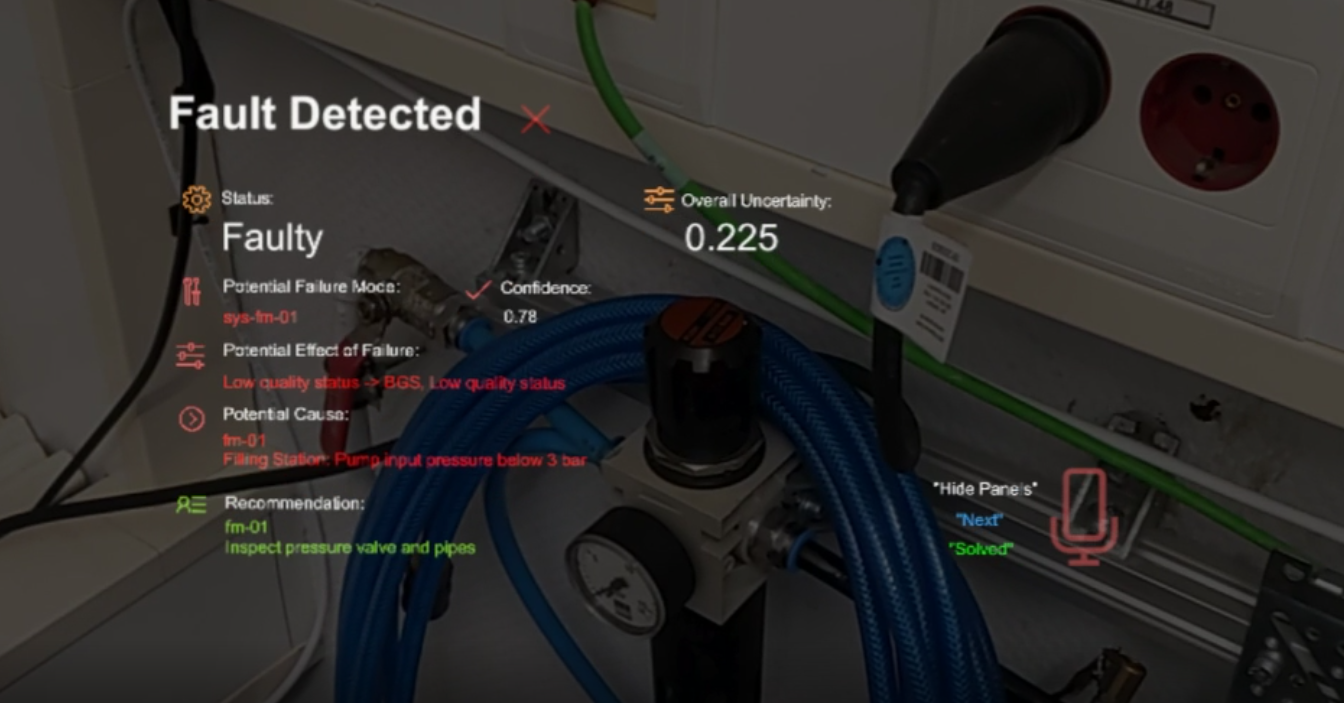}
    	\caption{Fault detection}\label{figure__use_case_hololens__2}
	\end{subfigure}
	\begin{subfigure}[b]{0.45\textwidth}
    	\includegraphics[width=\textwidth,keepaspectratio]{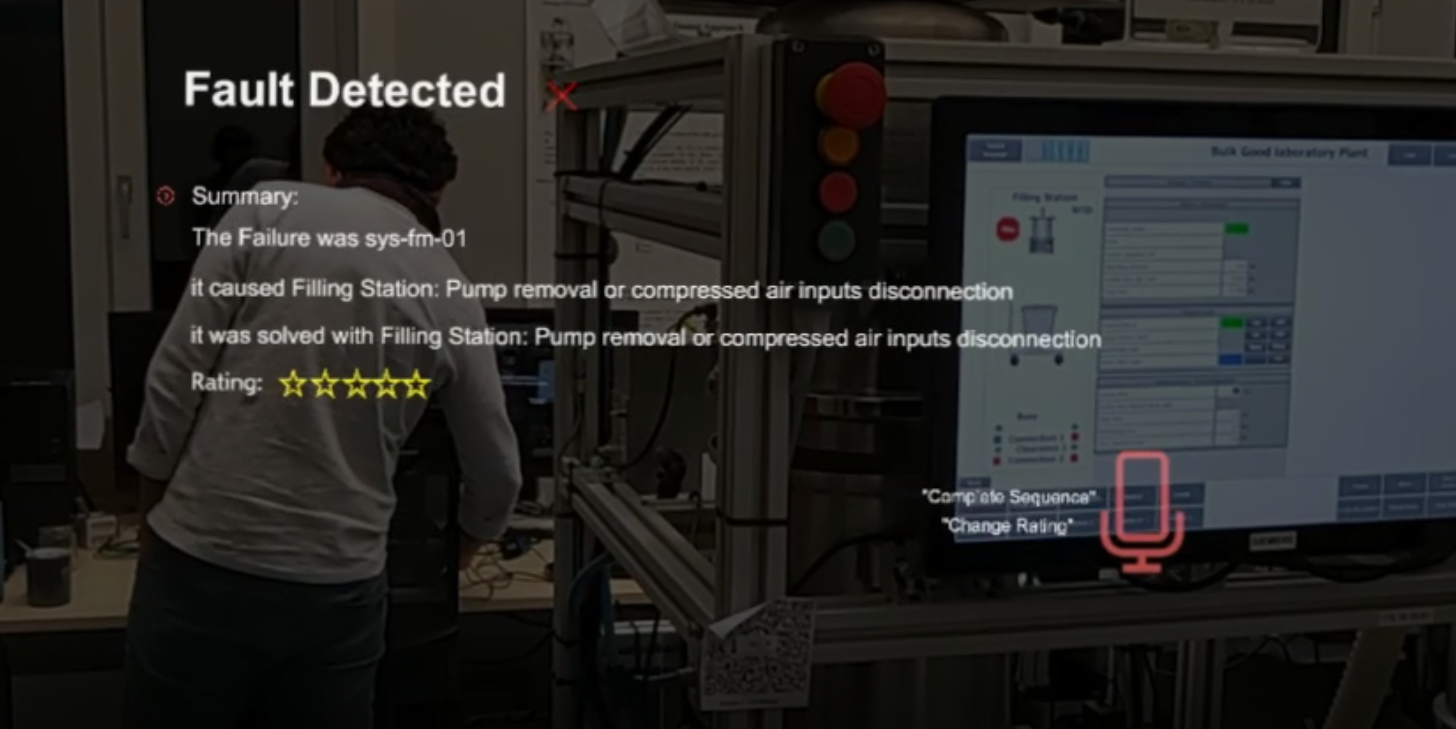}
    	\caption{User rating}\label{figure__use_case_hololens__3}
	\end{subfigure}
	~
	\begin{subfigure}[b]{0.45\textwidth}
    	\includegraphics[width=\textwidth,keepaspectratio]{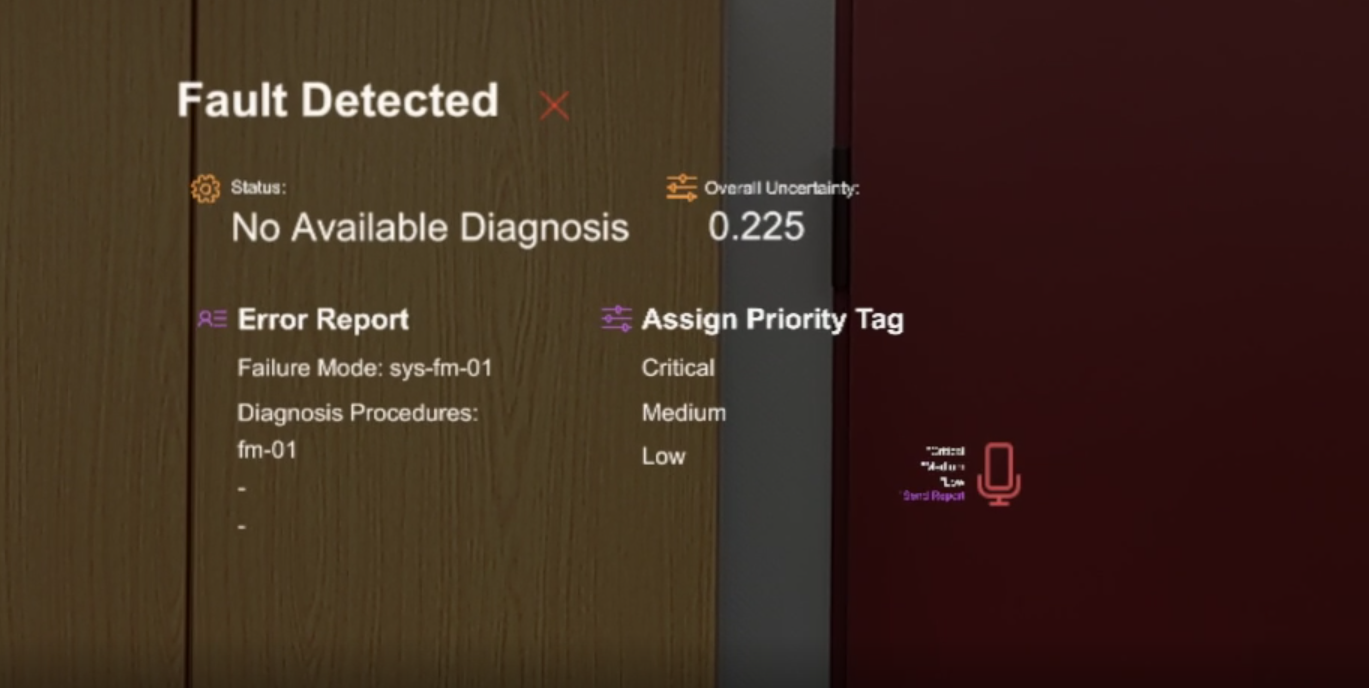}
    	\caption{No diagnosis and report}\label{figure__use_case_hololens__4}	
	\end{subfigure}
	\caption{Interactive User Assessment using Augmented Reality}\label{figure__use_case_hololens__5}
\end{figure*}

The backend collects data from the background and evaluates knowledge rules. HoloLens runs an internal loop and remains on standby until it receives a message from the backend.
The scenario started with an inexperienced operator wearing HoloLens. The BGS operates in a normal condition, and thus the HoloLens displays \textit{no fault}, see Fig. \ref{figure__use_case_hololens__1}.

The compressed air pressure valve is closed, which triggers a system failure mode with \textit{low quality} at the backend. The backend sends an assessment message to HoloLens containing the following information regarding the failure mode (e.g., description, effect, causes, and recommendations). In addition, it contains the weight of the failure mode and overall uncertainty of the system. This uncertainty was calculated by transforming the active failure mode into a set of evidence. HoloLens displays the assessment message, as shown in Figure \ref{figure__use_case_hololens__2}. 
The causes and recommendations of the system failure mode are the associated failure modes at the component level. Thus, a system failure mode triggers more than one component failure mode. The operator uses the voice command to select the next cause/recommendation pair, saying \textit{Next} (e.g., in the case of
several causes/recommendations) or \textit{solved} (e.g., in the case where the failure mode has been addressed). 
After the failure mode has been categorized as \textit{solved}, HoloLens displays a summary of the current failure mode and requests a \textit{user rating} to rate the user satisfaction from one to five stars. In the case where there is no available diagnosis, which means that the HoloLens reached the last pair of causes/recommendations, the HoloLens requests a report from the user, as shown in Figure \ref{figure__use_case_hololens__4}.
As displayed on the screen, the HoloLens requests an error report from the user. The backend receives a message with either the \textit{solved} or \textit{report} status, and it assigns the \textit{KPI compliance} as 1.0, or 0.0, respectively. The backend updates the weight of failure mode $w_{R}$ using the \textit{expert panel} weight $w_{P}$, \textit{KPI compliance} weight $w_{K}$, and \textit{user rating} weight $w_{U}$. 

\subsection{Discussion}
KLAFATE presented a way to formalize tacit knowledge and integrate it into an interactive assessment system. Remarkable features are the uncertainty representation of knowledge, validation of knowledge rules, and implementation of the framework into a small-scale industrial testbed. The data collection module allowed us to quantify knowledge uncertainty and validate new process recipes.
The limitations of this approach include multi-fault scenarios at the system level. Currently, only mutually exclusive faults are explicitly addressed. Addressing a scenario with simultaneous faults requires special treatment based on evidence theory, in which a combination of faults is considered. Consequently, the focal elements of the evidence increase to $2^{Faults}$.  
The knowledge model does not consider the historical nature of the fault, which means that the model cannot handle time-series data. This scenario can be addressed using a hybrid system composed of a current knowledge model and a machine learning model trained with time series.
Uncertainty quantification is based on the criteria given by the expert panel's weights, KPI analysis and user ratings. This uncertainty assigns a confidence level to the triggered operational rule of the knowledge model. 
Knowledge validation was performed using KPI compliance, specifically production rate. A typical industrial process include several KPIs to validate the process recipes (e.g., delay, machine availability, quality, and energy consumption). The methodology of KLAFATE and its implementation opens a discussion on the importance of user-centered approaches, especially in knowledge transfer and knowledge applicability on the shop floor.
 \section{Conclusions}\label{section_conclusions}
This research demonstrates how an interactive knowledge transfer framework can support the task of transforming tacit knowledge into explicit knowledge. The knowledge-based model was the outcome of this transformation, and was integrated into an interactive assistance system that could support the operator on the shop floor. In addition,  DSET quantified the uncertainty of the acquired knowledge, which was visually reflected in the results. The knowledge transfer framework provided a clear methodology for integrating uncertainty with the rules generated for the knowledge-based model.
The findings of this research would stimulate the discussion of how to transfer knowledge from the shop floor into a more institutionalized version, specifically, as a knowledge-based model embedded into an interactive assistance system. Furthermore, this novel methodology extracts expert domain knowledge that can be widely used in other disciplines that rely on expert feedback. The use of the DSET presents a new method for quantifying the uncertainty of expert knowledge. The integration of DSET and the knowledge-based model provides more reliable support to the operator, as it provides the assessment with a degree of certainty, which means that the operator can still use her own expertise to make the final decision.
The uncertainty plot helps in the decision-making process when validating a new body of knowledge, specifically when adopting a new recipe or set of setpoints. The validation plot portrayed KPI behavior while using a new body of knowledge, specifically new recipes (e.g., bad recipes yielded low KPIs, which consequently led to discarding the new recipe, whereas high KPIs encouraged the adoption of the recipe). The KLAFATE application presented an early adoption of the knowledge framework in an industrial setup. The demonstration provided a detailed sequence of the steps to be followed, as well as the results obtained after each step.

Although the present study provided a holistic approach to managing the knowledge chain through an interactive knowledge transfer framework, new questions arose during the development of this research. These new questions rely on the human nature of the information, specifically intrinsic bias, while extracting knowledge. This bias can play a significant role during the selection of knowledge to be included in the model and during the selection of criteria to quantify the uncertainty. These limitations must be addressed to adopt the knowledge framework into a fully automatic scenario. To this end, further research could explore new knowledge extraction strategies and methodologies to quantify uncertainty. Finally, knowledge internalization is a prospective line of research that should address the internalization institutionally and at the operator level. 

\bibliographystyle{IEEEtran}
\bibliography{IEEEabrv,ref}

\begin{IEEEbiography}[{\includegraphics[width=1in,height=1.25in,clip,keepaspectratio]{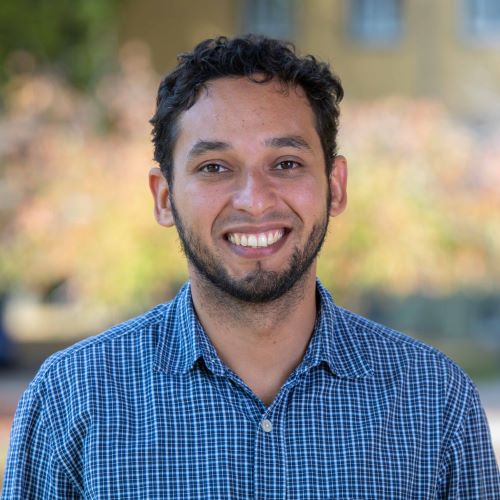}}]{Fernando Ar\'{e}valo N.} 
received his Engineer degree in electrical engineering from the Universidad de El Salvador, San Salvador, El Salvador, in 2005. He received his M.Sc. degree in systems engineering and engineering management from the South Westphalia University of Applied Sciences, Soest, Germany, in 2012. He is currently working toward the Ph.D. degree at the Ruhr-Universit\"{a}t Bochum, Bochum, Germany.
From 2008 to 2015, he worked as Project Engineer for Kimberly Clark, MTU Friedrichshafen, among others. From 2016 to 2021, he worked as a research assistant for the automation technology department in the South Westphalia University of Applied Sciences. Currently, he works as a data scientist for the company com2m GmbH.
His research interests include data-driven fault diagnosis, information fusion based on uncertainty, knowledge extraction, and application in the process industry and manufacturing. 
\end{IEEEbiography}

\begin{IEEEbiography}[{\includegraphics[width=1in,height=1.25in,clip,keepaspectratio]{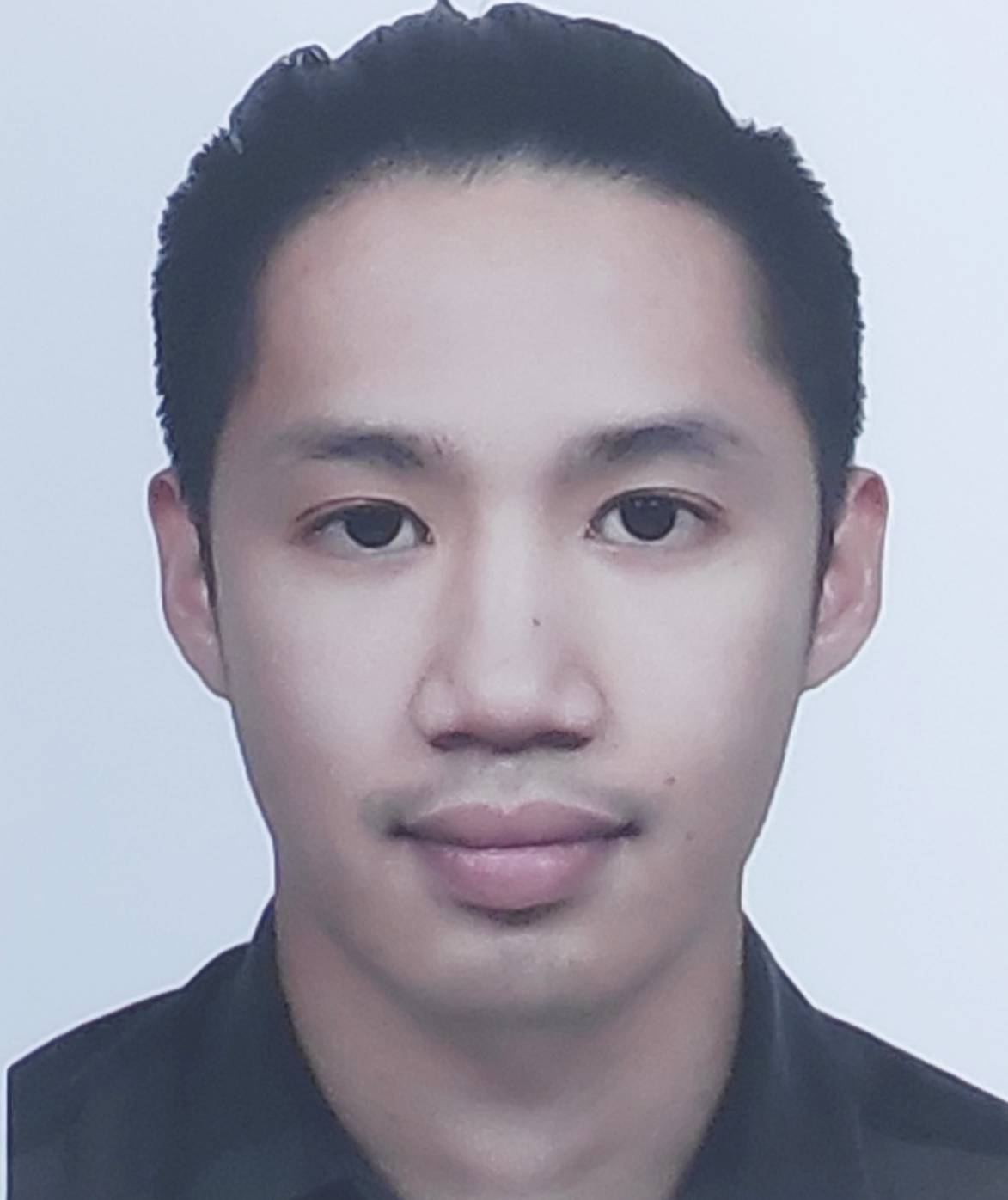}}]{Christian Alison M. P.} 
received his Bachelor dual-degree in industrial engineering from the Swiss German University, Indonesia, and South Westphalia University of Applied Sciences, Soest, Germany in 2020. He is currently working towards his M.Sc. degree in systems engineering and engineering management at the South Westphalia University of Applied Sciences, Soest, Germany.
His research interests include the future applications of augmented reality for process improvement and knowledge internalization.
\end{IEEEbiography}

\begin{IEEEbiography}[{\includegraphics[width=1in,height=1.25in,clip,keepaspectratio]{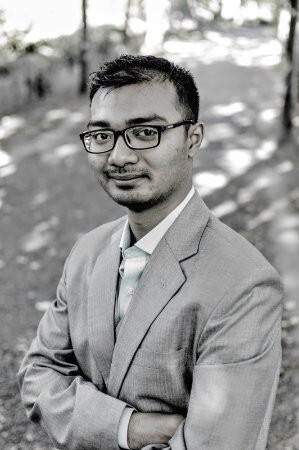}}]{M. Tahasanul Ibrahim} 
received his Bachelor degree in electrical and electronics engineering from the American International University, Bangladesh, in 2014. He received his M.Sc. degree in systems engineering and engineering management from the South Westphalia University of Applied Sciences, Soest, Germany, in 2020.
Since 2021 he works as a research assistant for the automation technology department in the South Westphalia University of Applied Sciences.
His research interests include information fusion based on uncertainty, rule-based systems, and back-end development.
\end{IEEEbiography}


\begin{IEEEbiography}[{\includegraphics[width=1in,height=1.25in,clip,keepaspectratio]{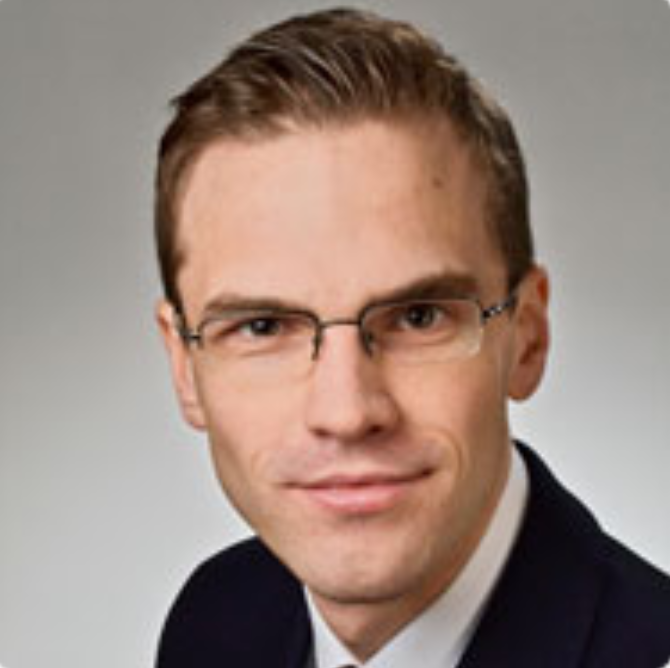}}]{Andreas Schwung} 
received the Ph.D. degree in electrical engineering from the Technische Universit\"{a}t Darmstadt, Darmstadt, Germany in 2011.
From 2011 to 2015, he was an R\&D Engineeer with MAN Diesel \& Turbo SE, Oberhausen, Germany. Since 2015, he has been a Professor of automation technology at the South Westphalia University of Applied Sciences, Soest, Germany. 
His research interests include model-based control, networked automation systems, and intelligent data analytics with application in manufacturing, process industry, and electromobility.
\end{IEEEbiography}


\end{document}